\documentclass[letterpaper,onecolumn]{IEEEtran}

\usepackage{graphicx}
\usepackage{epstopdf}
\usepackage{amsmath}
\usepackage{amsfonts}
\usepackage{amssymb}
\usepackage{cite}
\usepackage{hyperref}
\usepackage[justification=centering, font=normalsize]{caption}

\begin{document}

\title{Data-Driven Modeling of Noise Time Series with Convolutional Generative Adversarial Networks}

\author{Adam Wunderlich and Jack Sklar
\thanks{A. Wunderlich and J. Sklar are with the Communications Technology Laboratory, National Institute of Standards and Technology, Boulder, Colorado, 80305 USA. e-mail: adam.wunderlich@nist.gov. U.S. government work not protected by U.S. copyright.}}

\maketitle

\begin{abstract}
Random noise arising from physical processes is an inherent characteristic of measurements and a limiting factor for most signal processing and data analysis tasks.  Given the recent interest in generative adversarial networks (GANs) for data-driven modeling, it is important to determine to what extent GANs can faithfully reproduce noise in target data sets.  In this paper, we present an empirical investigation that aims to shed light on this issue for time series.  Namely, we assess two general-purpose GANs for time series that are based on the popular deep convolutional GAN (DCGAN) architecture, a direct time-series model and an image-based model that uses a short-time Fourier transform (STFT) data representation.  The GAN models are trained and quantitatively evaluated using distributions of simulated noise time series with known ground-truth parameters.  Target time series distributions include a broad range of noise types commonly encountered in physical measurements, electronics, and communication systems: band-limited thermal noise, power law noise, shot noise, and impulsive noise.  We find that GANs are capable of learning many noise types, although they predictably struggle when the GAN architecture is not well suited to some aspects of the noise, e.g., impulsive time-series with extreme outliers.  Our findings provide insights into the capabilities and potential limitations of current approaches to time-series GANs and highlight areas for further research.  In addition, our battery of tests provides a useful benchmark to aid the development of deep generative models for time series. 
\end{abstract}


\section{Introduction}
Noise, commonly defined as an unwanted, irregular disturbance, is a fundamental aspect of all real-world signals that arises from a multitude of natural and man-made sources.  Furthermore, the unpredictable, stochastic nature of noise makes it a significant impediment to measurement, data analysis, and signal processing.  Efforts to understand, mitigate, and harness the effects of noise over the last century have led to the extensive development of many physical and mathematical models, e.g., see \cite{Vasilescu2006,Milotti2019,Barrett2004,Howard2016} for overviews.

With recent advances in computational hardware and measurement equipment, it is now possible to collect, store, process, and analyze much larger quantities of data than previously possible.  Consequently, flexible, data-driven methods for signal modeling and processing are increasingly becoming feasible in many areas of science and engineering \cite{Montans2019}.  One form of data-driven modeling that has rapidly progressed in recent years is deep generative modeling, a type of unsupervised machine learning that uses deep neural networks to implicitly represent complicated, high-dimensional data distributions defined by a target training set \cite{Foster2019,Langr2019,Bond2021,Ruthotto2021}.  Deep generative models, and most notably, generative adversarial networks (GANs), have been used to successfully synthesize highly-realistic images, audio, video, and text \cite{Creswell2018, Hong2019, Wang2019, Pan2019}.  Moreover, because deep generative models can learn unknown, unstructured high-dimensional target distributions, they represent a potentially powerful class of methods for many data analysis and signal processing problems \cite{Foster2019,Langr2019,Bond2021,Ruthotto2021}. 

Since noise is a key component of realistic signals, and given the flourishing interest in GANs in particular, it is important to ask: To what extent are GANs capable of learning various noise types?  Answering this question provides insight into potential applications and limitations of GANs and related generative models.  

Prior related works includes the application of GANs to image denoising \cite{Chen2018,Wolterink2017,Lin2019,Ma2020,Cha2021}, image noise adaption \cite{Zhang2020,Miller2021}, image texture synthesis \cite{Zhou2018,Mauduit2020,Baradad2021}, and underwater acoustic noise modeling \cite{Zhou2021}.  Although the aforementioned investigations provide some evidence that GANs can learn noise and related distributions, they are limited to particular classes of noise and domain-specific applications.  Further, because most prior related studies focus on images, they give little insight into time series, which are of primary interest in many domains.

The literature on GANs for time series has predominantly focused on audio applications.  Many time-series GAN models leverage prior work on GANs for images by training the generator to produce an image-domain, time-frequency representation, such as a spectrogram, which is then mapped into a time series, e.g., \cite{Donahue2019, Kumar2019, Smith2020, Engel2019, Marafioti2019, Nistal2021}.  Additionally, there has been some work on GANs that directly model time series, e.g., using recurrent neural networks \cite{Esteban2017, Yoon2019}, or convolutional neural networks \cite{Donahue2019, Wiese2020}.

In the present work, we empirically investigate the ability of general-purpose GANs for time series to learn noise modeled as a real-valued, discrete-time random process.  Namely, as outlined in Section~\ref{sec:noise_models}, we examine four wide-ranging classes of noise commonly encountered in physical measurement, electronics, and communication:  band-limited thermal noise, power law noise, shot noise, and impulsive noise.  Within each noise class, we consider multiple random process models over a broad range of parameter values.  The mathematical noise models that we consider include stationary, nonstationary, Gaussian, non-Gaussian, and long-memory random processes.

Our evaluations focus on two complementary GAN models for time series based on the popular deep convolutional generative adversarial network (DCGAN) \cite{Radford2015} architecture: a direct time-series model, WaveGAN \cite{Donahue2019}, and an image-domain model that uses a complex-valued, short-time Fourier transform (STFT) representation of the time-series \cite{Engel2019,Nistal2021}.  Details are provided in Section~\ref{sec:GAN_models}.  The GAN architectures were selected for their general-purpose nature, relative simplicity, and straightforward implementation.  A prior investigation assessed the effectiveness of these GAN models for synthetic baseband communication signals in the presence of additive white noise and signal distortions arising from stochastic communication channels \cite{Sklar2021}.

Given the extraordinary number and breadth of noise models \cite{Vasilescu2006,Milotti2019,Barrett2004,Howard2016} and GAN architectures \cite{Creswell2018, Hong2019, Wang2019, Pan2019}, it is not feasible to examine all possibilities, and our investigations are necessarily limited in scope.  In particular, we do not aim to comprehensively evaluate all published GAN models for time series or to propose a single GAN architecture that works optimally for all noise types.  Instead, our goal is to assess the effectiveness of simple, general-purpose, convolutional GAN models for time series.  Nonetheless, to our knowledge, this investigation is the most extensive appraisal of GAN performance across a wide range of noise models thus far.       

Our empirical studies yield new insights into the capabilities and potential limitations of current approaches to time-series GANs and highlight areas for further research.  In addition, our battery of tests provides a useful benchmark to aid future developments.  Python software implementing our experiments and evaluations as well as training datasets and results are publicly available \cite{noiseGAN-software, noiseGAN-data}.   

\section{Noise Models and Simulation Methods}
\label{sec:noise_models}

In this section, we review the classical noise models and the simulation methods used to generate target distributions for our experiments.  Specific parameter choices for our synthetic noise data sets are given in Section~\ref{sec:experiments}.

The mathematical models presented here were selected because they cover disparate, well-known noise types and because there are accurate, computationally efficient methods for simulation and parameter estimation.  This set of noise models is not comprehensive, and descriptions of additional types of noise can be found in \cite{Vasilescu2006,Milotti2019,Barrett2004,Howard2016}.  For simplicity, we focus on real-valued random processes.  

Below, using standard notation, we denote the set of real numbers as $\mathbb{R}$ and the set of integers as $\mathbb{Z}$.  All simulated time series were $4096$ samples long, which provided a good balance between realism and computational complexity.  Throughout, unitless quantities, e.g., time, are used.

\subsection{Band-Limited Thermal Noise}
Thermal noise, also called Johnson-Nyquist noise, arises from the thermal motion of charge carriers inside an electrical conductor \cite{Vasilescu2006,Milotti2019,Barrett2004}.  Thermal noise is commonly modeled as a zero-mean white process, i.e., a sequence of independent, identically distributed (i.i.d.) random variables with zero mean and finite variance \cite{Howard2016,Percival2000}.  In the case of radio frequency electronics, thermal noise is band-limited by system components.  For this reason, band-limited (or filtered) thermal noise is of interest in many contexts \cite{Papoulis1965,Howard2016}.

\begin{figure}[htb!]
\centering
\includegraphics[width=0.4\linewidth]{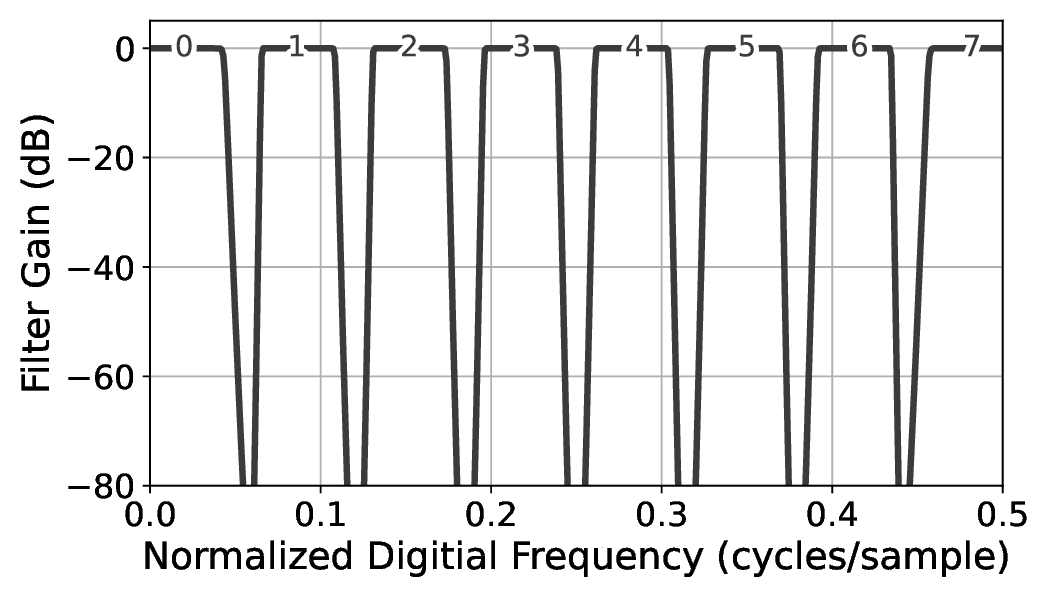} 
\caption{Frequency response of each digital filter used to simulate band-limited thermal noise, indexed 0 through 7.}
\label{fig:BPfilter_responses}
\end{figure}

To simulate band-limited thermal noise, we first generated a white standard normal sequence and then filtered it with a digital bandpass filter.  Specifically, we applied a 40th order digital Butterworth filter, implemented using cascaded second-order sections and zero-phase filtering \cite{Mitra2001}.  Frequency responses for the eight bandpass filters used to generate our target distributions are shown in Figure~\ref{fig:BPfilter_responses}.

\subsection{Power Law Noise}
\label{sec:power_law_noise_models}

Power law noise, also called colored, fractional, or fractal noise, arises in electronics as well as a diverse array of other physical phenomena \cite{Milotti2019,Barrett2004,Keshner1982,Vasilescu2006}.  Power law noise noise is characterized by a power spectral density (PSD), $S(f)$, that is proportional to a power of frequency, $f$, at low frequencies, i.e., $S(f) \propto |f|^{\eta}$, where $\eta$ is a real number.  Specific integer powers are often associated with ``colors of noise'', e.g., random processes with $\eta=-1$, $0$, and $1$ are called pink, white, and blue noise, respectively \cite[Ch.~3]{Rouphael2014}.  When $\eta$ is near $-1$, the process is also known as $1/f$ noise, flicker noise, or excess noise \cite{Keshner1982,Milotti2019}.  

We consider two well-known mathematical models for power law noise: fractional Gaussian noise (FGN) and fractional Brownian motion (FBM) \cite{Mandelbrot1968}.  FGN can be interpreted as a generalization of discrete-time white Gaussian noise, and FBM can be interpreted as a generalization of the continuous-time Brownian motion (or Wiener) process \cite{Mandelbrot1968}.  The above models arise in the study of self-similar (or fractal) processes, as well as so-called long memory (or persistent) processes \cite{Beran2013,Percival2000,Veenstra2013}.

FGN is a zero-mean, stationary, discrete-time Gaussian process, $Y_t$, $t\in \mathbb{Z}$, with autocovariance sequence
\begin{align}
    \text{Cov}(Y_t,Y_{t+k}) = \frac{\sigma_Y^2}{2}\left( |k-1|^{2H} + |k+1|^{2H} - 2|k|^{2H} \right),
\end{align}
where $k\in\mathbb{Z}$, $\sigma_Y^2 = \text{Var}[Y_t]$, and $H\in (0,1)$ is called the Hurst index \cite{Mandelbrot1968,Percival2000,Beran2013}.  Successive time steps of FGN are negatively correlated when $H \in (0,0.5)$ and positively correlated when $H \in (0.5,1)$.  When $H=0.5$, successive time steps are uncorrelated, and FGN reduces to classical white Gaussian noise.  Near $f=0$, the PSD for FGN is proportional to $|f|^{1-2H}$ \cite{Mandelbrot1968,Percival2000,Veenstra2013}.

FGN arises as the increment process of the continuous-time FBM process, $B_H(t), t \in \mathbb{R}$ \cite{Mandelbrot1968,Percival2000,Beran2013}.  Namely,
\begin{equation}
   Y_t = B_H(t+1) - B_H(t), \quad t \in \mathbb{Z}.
\end{equation}
For the rigorous definition of FBM, see, e.g., \cite{Mandelbrot1968,Beran2013}.  FBM is a nonstationary, zero-mean Gaussian process with $B_H(0)=0$, and in the special case $H=0.5$, FBM reduces to a classical Brownian motion process.  The PSD for the nonstationary FBM process can be given a rigorous interpretation \cite{Flandrin1989}, where the PSD is proportional to $|f|^{-(2H+1)}$ \cite{Mandelbrot1968,Flandrin1989,Percival2000}.  

In summary, as $f \rightarrow 0$, the PSDs for FGN and FBM are proportional to $|f|^\eta$, where $\eta \in (-1,1)$ for FGN and $\eta \in (-3,-1)$ for FBM, respectively, with $H\in (0,1)$.

\begin{figure}[htb!]
\centering
\includegraphics[width=0.49\linewidth]{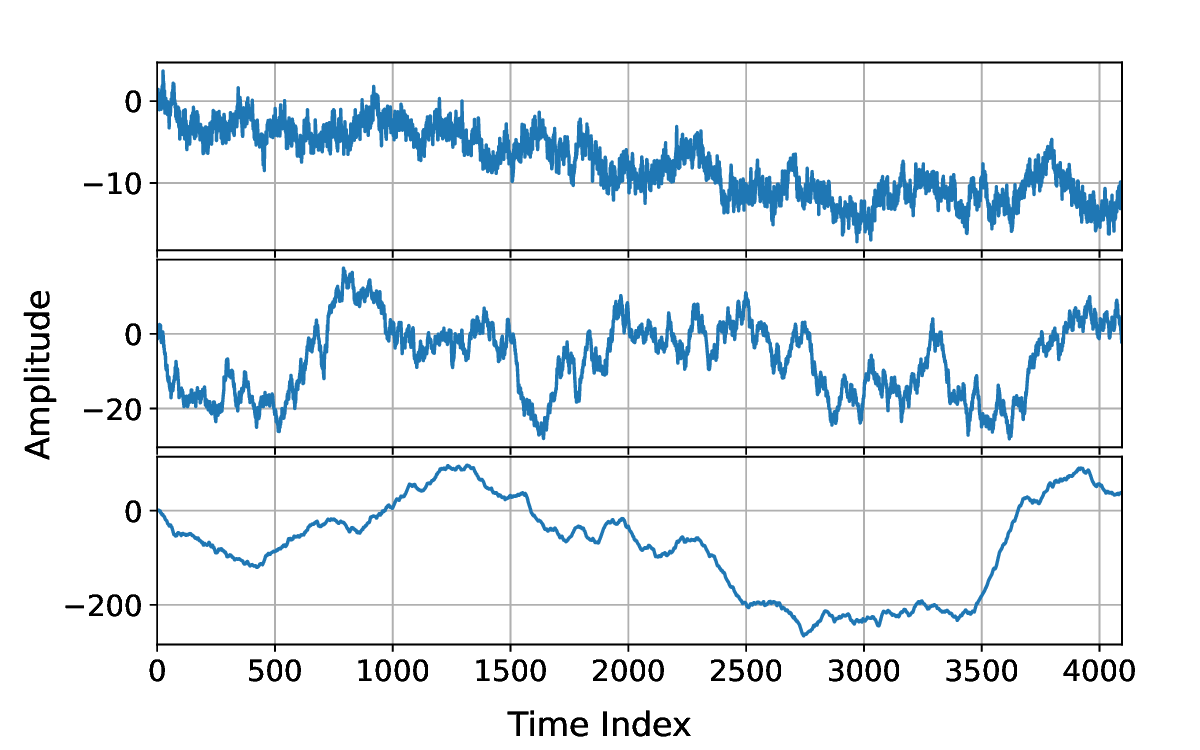} 
\includegraphics[width=0.49\linewidth]{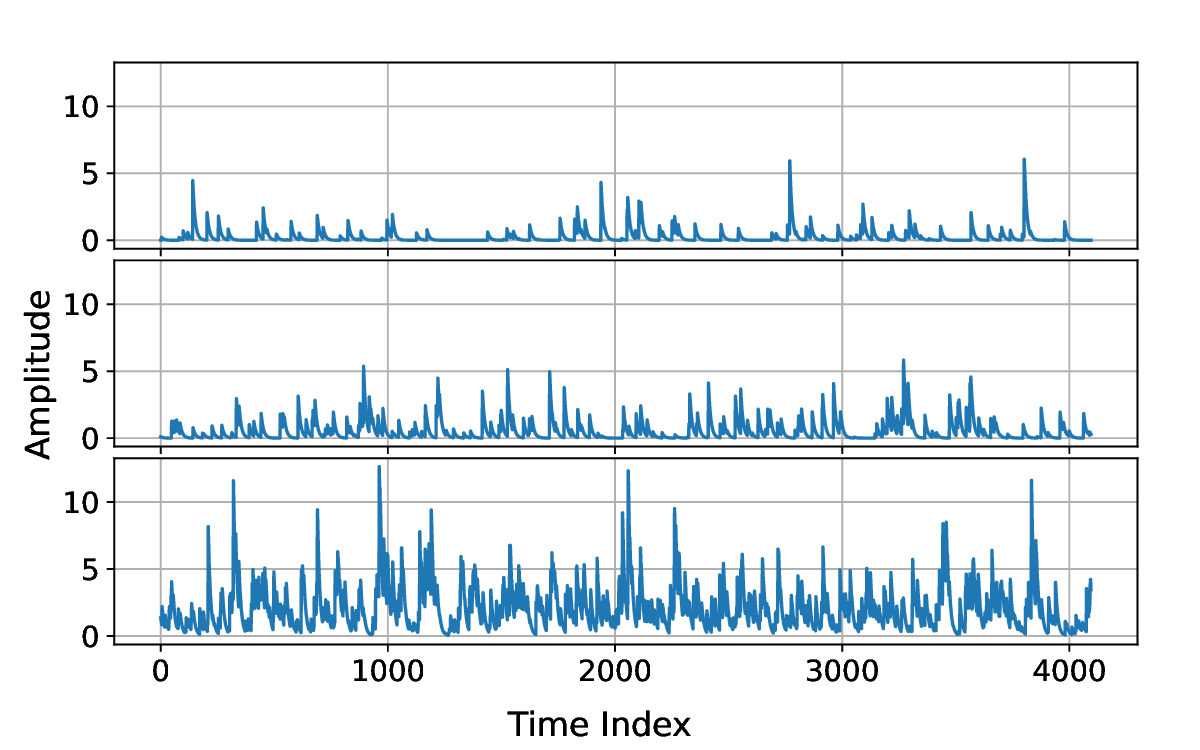} 
\caption{Left: Example fractional Brownian motion time series with $H=0.2$, $0.5$, and $0.8$ from top to bottom.  Right: Example shot noise time series with one-sided exponential pulse type and event rates $\nu=0.25$, $0.5$, and $2$ from top to bottom.}
\label{fig:FBM_shot_examples}
\end{figure}

To synthesize discrete-time FGN and FBM, we implemented the exact approach of Perrin et al. \cite{Perrin2002}, which utilizes the fast circulant embedding method \cite{Dietrich1997} to generate FGN and applies cumulative summation to obtain discrete-time FBM.  All FGN simulations set $\sigma_Y^2=1$.  Examples of synthetic FBM time series are shown in Figure~\ref{fig:FBM_shot_examples} (Left) for $H=0.2$, $0.5$, and $0.8$.  It can be seen that as the Hurst index increases, FBM tends to deviate further from the origin.  

\subsection{Shot Noise}
\label{sec:shot_noise_model}

Shot noise, also called Poisson noise or photon noise, arises from the random arrival of discrete charge carriers in electronics and photons in optics \cite{Howard2016,Barrett2004,Vasilescu2006,Milotti2019}.  Shot noise can be modeled using a filtered Poisson process of the form
\begin{equation}
    X(t) = \sum_{n=1}^{N(t)} A_n p(t-\tau_n), \quad t>0,
\label{eq:shot_noise_model}
\end{equation}
where $N(t)$, the number of events in the interval $(0,t]$, is a homogeneous Poisson point process with event rate $\nu$ and event times, $\{\tau_n\}$ \cite{Parzen1962,Snyder1991,Barrett2004,Howard2016,Theodorsen2017}.  If $N(t)=0$, then the sum is taken to be zero.  Above, $p(t)$ is a deterministic pulse function, and the pulse amplitudes, $\{A_n\}$, are independent, identically distributed, and independent of $N(t)$.  For a finite time interval of length $T$, the number of events, $N$ is Poisson distributed with mean $\nu T$, and the event times, $\{\tau_1,\tau_2,\ldots,\tau_N\}$, are uniformly distributed on the interval \cite[p.~140]{Parzen1962}.  

Following Theodorsen et al. \cite{Theodorsen2017}, we assumed that the pulse amplitudes follow an exponential distribution with mean $\beta$.  We considered two pulse functions, one-sided exponential and Gaussian, taken from Howard \cite[p.~506]{Howard2016}. 
 Table~\ref{tab:pulse_functions} summarizes the pulse functions, where $u(t)$ denotes the unit step function equal to one for $t\geq 0$ and zero otherwise, and $\sigma_d$ is a pulse duration parameter.  Table~\ref{tab:pulse_functions} also lists the integrals $I_1 = \int_{-\infty}^{\infty} p(t)\,dt$ and $I_2 = \int_{-\infty}^{\infty} p(t)^2 \,dt$ of each pulse function, which are used for the event rate estimator introduced in Section~\ref{sec:noise_param_estimation}.  

For a finite time interval, $(0,T]$, the mean and autocovariance of the shot noise process are time-dependent, approaching steady-state values as $t, T \rightarrow \infty$ \cite{Howard2016}.  Therefore, to approximate a weak-sense stationary discrete-time shot noise process, we generated a length $2L$ process of duration $T=(2L-1)\Delta_t \gg \sigma_d$ and then discarded the first $L$ samples.  Namely, defining a discrete-time grid $t_m = m\Delta t$, for $m=0,1,\ldots,2L-1$, we drew $N$ from a Poisson distribution with mean $\nu T$, where $T = (2L-1)\Delta t$.  Next, we randomly generated $N$ integers $\{m_1, m_2, \ldots, m_N\}$ from a discrete uniform distribution on $[0, 2L-1]$ and drew $\{A_1,A_2,\ldots,A_N\}$ independently from an exponential distribution with mean $\beta$.  Then, we formed the impulse sequence
\begin{equation}
    f[m] = \sum_{n=1}^N A_n \delta_{m,m_n},
\end{equation}
where $\delta_{m,m_n}$ is a Kronecker delta function, and performed the discrete convolution of $f[m]$ with the sampled pulse function, $p(t_m)$, retaining the $2L$ samples in the middle of the convolution result.  Last, we discarded the first $L$ samples to remove any transients and approximate a steady-state realization of a length $L$ discrete-time shot noise process.  The validity of the steady-state simulated shot noise time series was verified by checking that there was close agreement between the empirical autocovariance function and the theoretical asymptotic autocovariance function \cite{Howard2016}.  For all shot noise simulations, we set  $\sigma_d = 1$, $\beta=1$, and $\Delta t = 0.1$.  Example synthetic shot noise time series with a one-sided exponential pulse function and event rates $\nu=0.25$, $0.5$, and $2$ are shown in Figure~\ref{fig:FBM_shot_examples} (Right).  

\begin{table}[tb]
    \centering
    \caption{Pulse functions used to simulate synthetic shot noise.}
    \begin{tabular}{|l|l|c|c|}
        \hline
         \textbf{Pulse Type} &\textbf{p(t)} &$\mathbf{I_1}$ &$\mathbf{I_2}$ \\ 
         \hline & & &\\[-1.0em] 
          One-sided Exponential   
          &$\frac{1}{\sigma_d}\exp[-t/\sigma_d]u(t)$ &1 &$\frac{1}{2\sigma_d}$ \\[1.5ex]
          Gaussian &$\frac{1}{\sigma_d\sqrt{2\pi}}\exp[-t^2/(2\sigma_d^2)]$ &1 &$\frac{1}{2\sigma_d\sqrt{\pi}}$ \\[0.5em] 
          \hline
    \end{tabular}
    \label{tab:pulse_functions}
\end{table}

\subsection{Impulsive Noise}
\label{sec:impulsive_noise_models}

Impulsive noise, consisting of random, large bursts of short duration arising from either naturally occurring or man-made sources, is a limiting factor for many communication scenarios \cite{Pighi2009,Tsihrintzis1995,Ghosh1996,Herath2012}, including wireless \cite{Blackard1993,Mirahmadi2013}, digital subscriber line \cite{Kerpez1995,Mann2002}, power line \cite{Meng2005,Fernandes2017}, and undersea acoustic environments \cite{Kuai2016,Wang2020}.  Many models for impulsive noise have been developed; see Shongwe et al. \cite{Shongwe2014} for an overview.  We focused on two well-studied impulsive noise models that were straightforward to implement and evaluate: the Bernoulli-Gaussian and symmetric alpha-stable models, described below.  These models both define non-Gaussian, memoryless, white processes with a power spectrum that is constant across all frequencies.  Impulse noise models with memory have also been proposed, e.g., see \cite{Shongwe2014,Fernandes2017}, but such models are outside the scope of the present study.   

A simple impulsive noise model that has been applied in many contexts is the Bernoulli-Gaussian (BG) model \cite{Pighi2009,Herath2012,Ghosh1996,Fernandes2017}, independently defined at each discrete time step as 
\begin{equation}
    X_{BG} = N_w + B N_i,
\label{BG_noise}
\end{equation}
where $N_w$ and $N_i$ are independent, zero-mean, normal random variables with variances $\sigma_w^2$ and $\sigma_i^2$, respectively, and $B$ is a Bernoulli random variable with mean $p$, i.e., the probability that $B=1$ is $p$, where $p$ is called the impulse probability.  Above, $N_w$ corresponds to a thermal noise background and $N_i$ is intermittent impulsive noise.  The probability density function (PDF) for $X_{BG}$ is the Gaussian mixture 
\begin{equation}
    f(x) = (1-p)\mathcal{N}(x;0,\sigma_w^2) + p\mathcal{N}(x;0,\sigma_w^2+\sigma_i^2)
\label{eq:BG_mixture}
\end{equation}
where $\mathcal{N}(x;\mu,\sigma^2)$ denotes the PDF for a normal distribution with mean $\mu$ and variance $\sigma^2$.  

\begin{figure}[tb!]
\centering
\includegraphics[width=0.49\linewidth]{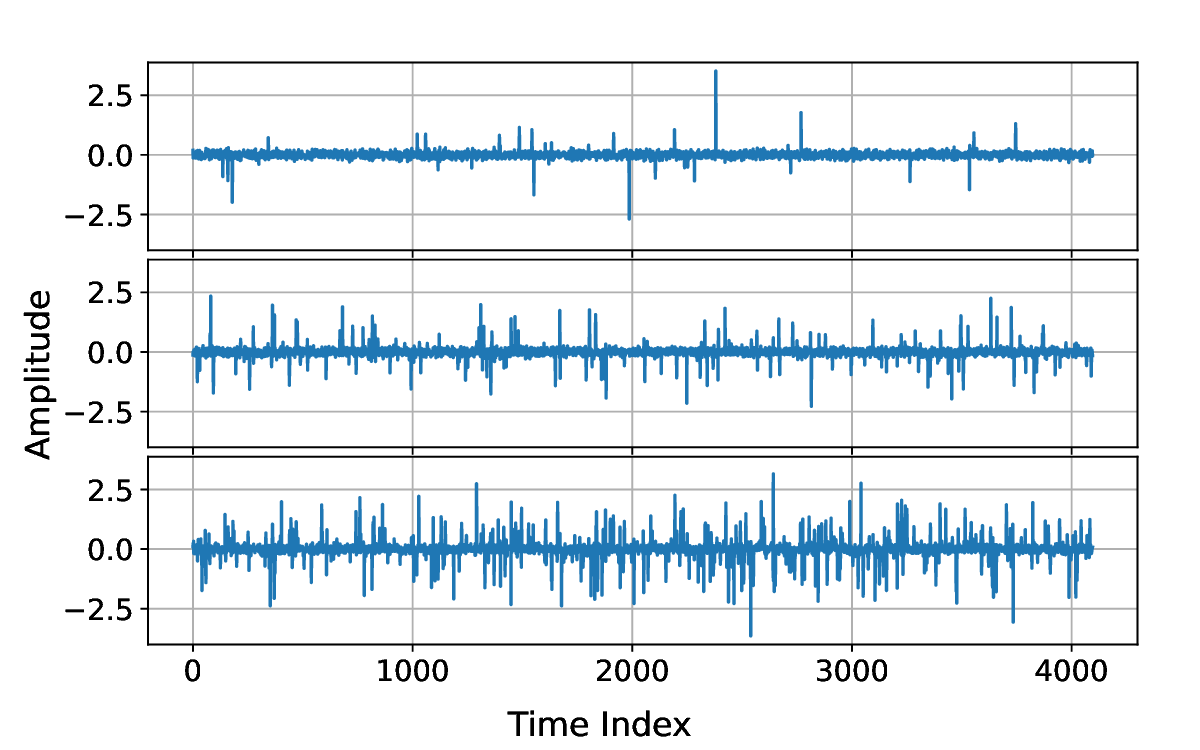} 
\includegraphics[width=0.49\linewidth]{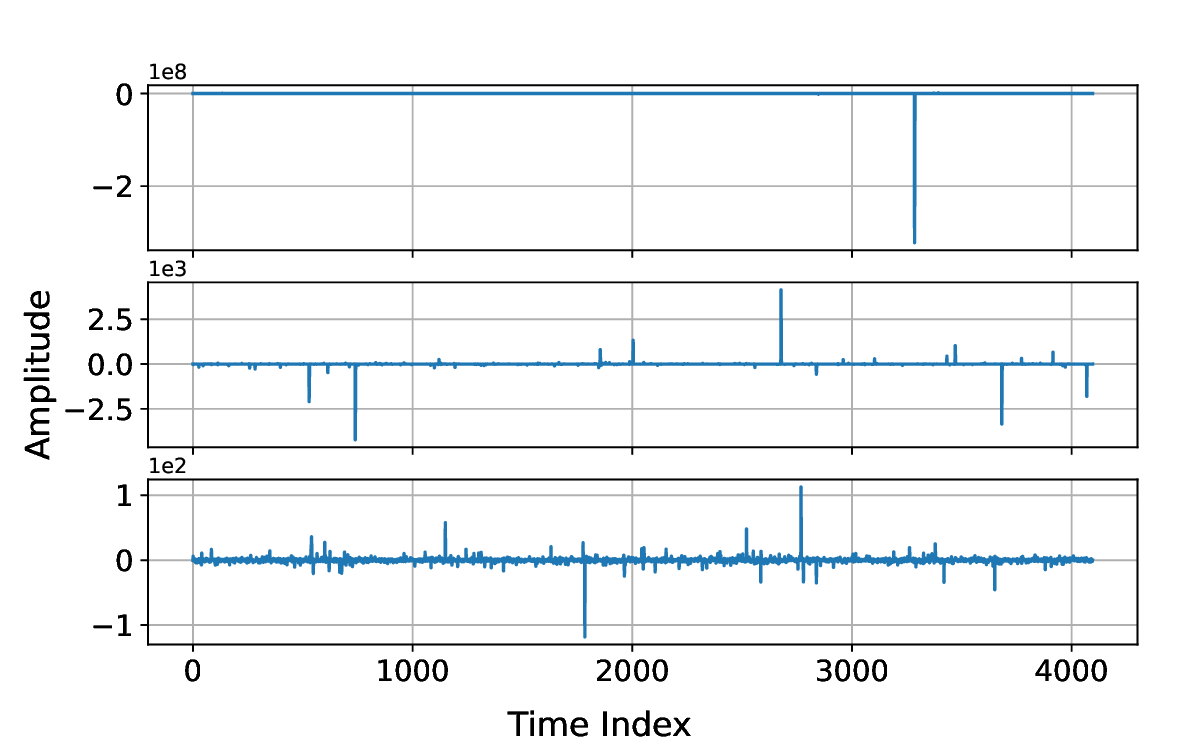} 
\caption{Example time series for impulsive noise.  Left: Bernoulli-Gaussian with $\sigma_w=0.1$ and $\sigma_i=1$ for $p=0.01$ (Top), $p=0.05$ (Middle), and $p=0.1$ (Bottom). Right: Standard symmetric $\alpha$-stable with $\alpha=0.5$ (Top), $\alpha=1$ (Middle), and $\alpha=1.5$ (Bottom)}
\label{fig:BG_SAS_examples}
\end{figure}

\begin{figure}[htb!]
\centering
\includegraphics[width=.3\linewidth]{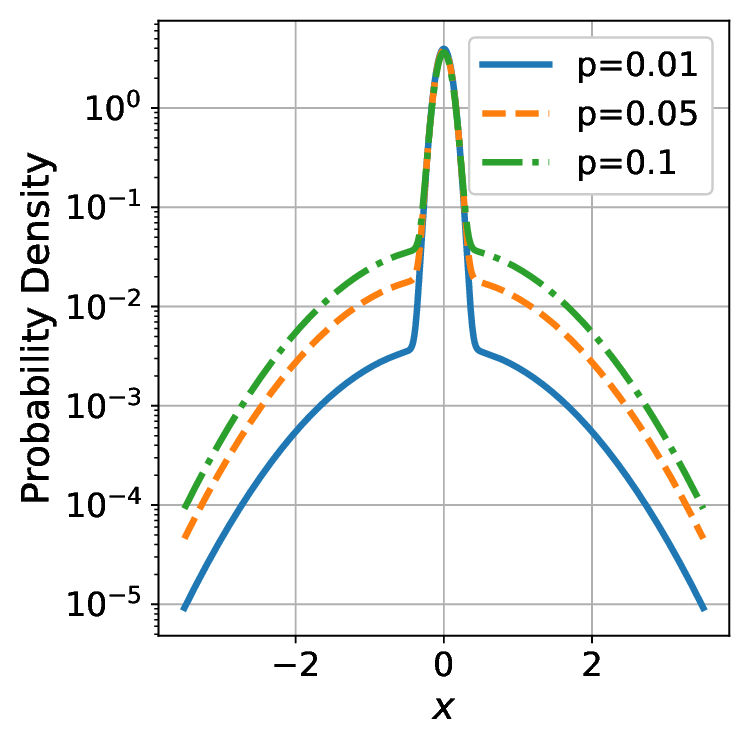} 
\includegraphics[width=.3\linewidth]{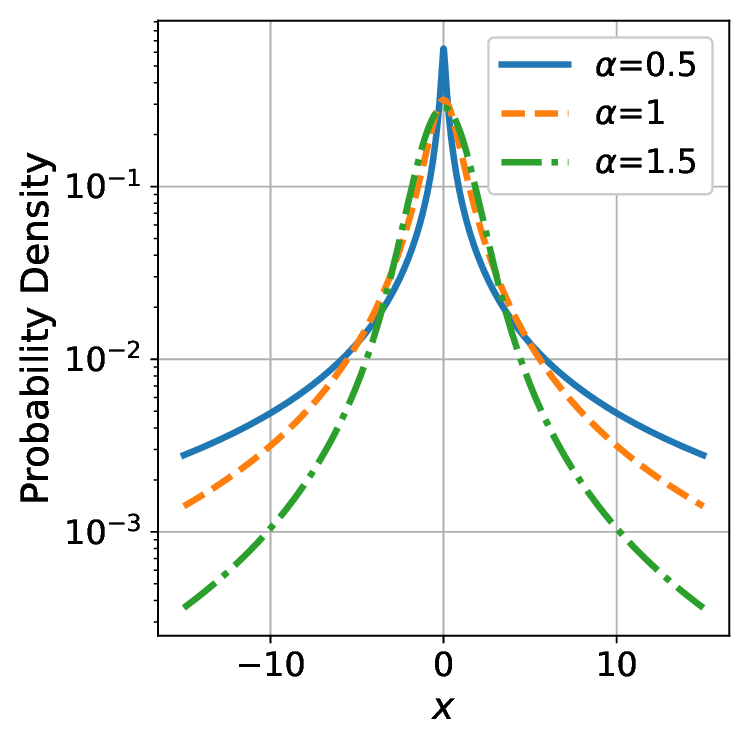} 
\caption{Example probability density functions for impulsive noise.  Left: Bernoulli-Gaussian with $\sigma_w=0.1$ and $\sigma_i=1$.  Right: standard symmetric $\alpha$-stable.}
\label{fig:impulse_noise_pdfs}
\end{figure}

We simulated independent BG noise at each time step using equation (\ref{BG_noise}) with $\sigma_w=0.1$ and $\sigma_i=1$.  Example time series are shown in Figure~\ref{fig:BG_SAS_examples} (Left) for $p=0.01$, $0.05$, and $0.1$.  Corresponding PDFs are plotted on a logarithmic scale on the left side of Figure~\ref{fig:impulse_noise_pdfs}.

Another popular model for impulsive noise is the symmetric $\alpha$-stable ($S\alpha S$) family of distributions,  a subclass of the stable (a.k.a. Levy $\alpha$-stable) family of distributions, which are used to model heavy-tailed, non-Gaussian phenomena \cite{Nolan2020,Nikias1995,Shao1993,Tsihrintzis1995,Georgiou1999}.  The PDF of a $S \alpha S$ distribution can be succinctly expressed in terms of its characteristic function as
\begin{equation}
    f(x;\alpha, \gamma, \delta) = \frac{1}{2\pi} \int_{-\infty}^{\infty} \exp\left(i\delta u -\gamma|u|^\alpha \right)e^{-iux} \,du,
\end{equation}
where $i^2=-1$, $\alpha \in (0,2]$ is the characteristic exponent, $\gamma>0$ is the scale parameter, and $\delta \in \mathbb{R}$ is the location parameter.\footnote{Many different parameterizations  for stable distributions are in use; see Nolan \cite[Secs. 1.3, 3.5]{Nolan2020} for an overview.  Here, we used the parameterization of Nikias and Shao \cite{Nikias1995}.}  A $S \alpha S$ distribution is said to be ``standard'' if $\delta=0$ and $\gamma=1$.  The special cases $\alpha = 1$ and $\alpha=2$ correspond to Cauchy and normal distributions, respectively.  As $\alpha$ decreases, the PDF has a sharper peak and the tails become heavier \cite{Nolan2020,Nikias1995}.  

We considered discrete time $S \alpha S$ processes where the value at each time step is an independent standard $S \alpha S$ random variable with parameter $\alpha$.  To simulate standard $S \alpha S$ variates, we used the `pylevy' Python module \cite{PyLevy2020}, which implements a method of Chambers et al. \cite{Chambers1976} for generating stable random variables; see also \cite{Nolan2020}.  Example time series are shown in Figure~\ref{fig:BG_SAS_examples} (Right) for three values of $\alpha$.  Corresponding PDFs are plotted on a logarithmic scale on the right side of Figure~\ref{fig:impulse_noise_pdfs}.

Comparing the example time series plots in Figure \ref{fig:BG_SAS_examples}, we see that the range of BG noise is fairly consistent across different impulse probabilities, $p$.  On the other hand, the range of $S \alpha S$ noise varies by several orders of magnitude for different values of the characteristic exponent, $\alpha$.  These observations are consistent with the corresponding PDFs shown in Figure~\ref{fig:impulse_noise_pdfs}.  Namely, because BG noise is a mixture of two Gaussian distributions, BG noise has rapidly decaying ``light'' tails, whereas $S\alpha S$ noise has slowly decaying ``heavy'' tails with a higher probability of extreme values \cite{Nair2022}.

\section{GAN Models}
\label{sec:GAN_models}

We implemented two CNN-based GAN models for our experiments that are based on the widely used DCGAN model \cite{Radford2015}: a 1-D convolutional model trained directly on time series, called WaveGAN \cite{Donahue2019} and a 2-D convolutional model trained on the complex-valued STFT.  Both models were designed to generate time series of length 4096.  Details on these models are given below.  We start with a brief introduction to GANs.  

\subsection{Basic GAN Theory}

Since the introduction of GANs 2014, research on GANs and related deep generative modeling frameworks has developed quickly and spawned a large literature.  For reviews, see \cite{Foster2019, Langr2019, Creswell2018, Hong2019, Wang2019, Pan2019, Bond2021, Ruthotto2021}.

Given a training set drawn from a high-dimensional target distribution, $p_d$, e.g., consisting of images or time series, the basic idea of a GAN is to train two deep neural networks, a generator network, $G$, and a discriminator network, $D$, together dynamically.  The generator generates samples from a generator distribution, $p_g$, where the aim is to match the target distribution.  The discriminator seeks to assess the realism of generated samples, i.e., determine if samples are `real' or `fake.'  The generator is fed a random vector, $z$, drawn from a specified latent distribution, e.g., multivariate uniform, which it maps to a generated sample, $G(z)$.  The discriminator maps sample data, $x$, to the probability that the sample belongs to the target distribution, $D(x)$.     

The generator and discriminator networks are typically trained using a backpropagation implementation of stochastic gradient descent with a specified loss, or objective function \cite{Goodfellow2016}.  Many different approaches to training GANs have been investigated to avoid failure modes such as inadequate convergence and mode collapse, where the generator output in insufficiently diverse.  The GAN models that we investigated were trained with the widely-used Wasserstein GAN loss with gradient penalty \cite{Gulrajani2017}, which seeks to minimize the Wasserstein distance between the generated distribution and the target distribution \cite{Arjovsky2017}.  Specifically, training aimed to minimize the objective function 
\begin{equation}
        \mathcal{L} = E_{\tilde{x}\sim p_g}[D(\tilde{x})] - E_{x\sim p_{d}}[D(x)] + \lambda~E[(\|\nabla_{\hat{x}}D(\hat{x})\|_2 - 1)^2],
\label{eq:WGP_loss}
\end{equation}
where $\hat{x} = \epsilon x + (1-\epsilon)\tilde{x}$ with $\tilde{x} \sim p_g$, $x \sim p_d$, and $\epsilon \sim U[0,1]$.  Here, $E[\cdot]$ denotes mathematical expectation, a tilde ($\sim$) indicates that a random variable follows a specified distribution, and $U[0,1]$ denotes the uniform distribution over the unit interval.  Following the implementation recommendations of Gulrajani et al. \cite{Gulrajani2017}, we set the gradient penalty coefficient, $\lambda$, equal to $10$.  Additional training details are provided in Section~\ref{sec:training}.    

\subsection{WaveGAN}

WaveGAN \cite{Donahue2019} is a direct time series GAN designed for audio generation based on a 1-D flattened version of the 2-D DCGAN model \cite{Radford2015}.  Tables~\ref{tab:WaveGAN-gen} and \ref{tab:WaveGAN-disc} outline our implementation of the WaveGAN generator and discriminator, respectively.  In these tables, Dense, Conv 1-D and Transpose Conv 1-D, denote dense fully connected layers, one-dimensional convolutional layers, and transposed convolutional layers, respectively. Also, Tanh, ReLU, and LReLU indicate hyperbolic-tangent (Tanh), rectified linear unit (ReLU), and leaky rectified linear unit (LReLU) activation functions.  The filter dimensions for convolutional layers correspond to kernel length, number of input channels, and number of output channels, respectively.  Similarly, the filter dimensions for the dense layers correspond to input length and output length, respectively.  The first output shape dimension, $n$, denotes the batch size.  Compared to the original WaveGAN model, which was designed to produce time series of length $16\,384$ our only modification was to change the dense layer to support the $4096$ length of our synthetic noise waveforms.

In the discriminator, WaveGAN includes an additional operation, called ``phase shuffle,'' consisting of a random circular shift on the activation output of each convolutional layer.  Our implementation applied a random circular shift between $-2$ and $2$ time steps, as recommended by Donahue \emph{et al.} \cite{Donahue2019}.

\begin{table}[p]
  \centering
    \caption{WaveGAN generator architecture.}
    \label{tab:WaveGAN-gen}
    \begin{tabular}{|l|c|r|} 
    \hline
      \textbf{Operation} & \textbf{Filter Shape} & \textbf{Output Shape}\\
      \hline
      $z$ $\sim$ Uniform(-1, 1) &  & ($n$, 100)\\
      Dense & (100, 4096) & ($n$, 4096)\\
      Reshape & & ($n$, 1024, 4)\\
      ReLU & & ($n$, 1024, 4)\\
      Transpose Conv1-D (stride=4) & (25, 1024, 512) & ($n$, 512, 16)\\
      ReLU & & ($n$, 512, 16)\\
      Transpose Conv1-D (stride=4) & (25, 512, 256) & ($n$, 256, 64)\\
      ReLU & & ($n$, 256, 64)\\
      Transpose Conv1-D (stride=4) & (25, 256, 128) & ($n$, 128, 256)\\
      ReLU & & ($n$, 128, 256)\\
      Transpose Conv1-D (stride=4) & (25, 128, 64) & ($n$, 64, 1024)\\
      ReLU & & ($n$, 64, 1024)\\
      Transpose Conv1-D (stride=4) & (25, 64, 2) & ($n$, 1, 4096)\\
      Tanh & & ($n$, 1, 4096) \\
      \hline
    \end{tabular}
\end{table}

\begin{table}[p]
  \centering
    \caption{WaveGAN discriminator architecture.}
    \label{tab:WaveGAN-disc}
    \begin{tabular}{|l|c|r|} 
    \hline
      \textbf{Operation} & \textbf{Filter Shape} & \textbf{Output Shape}\\
      \hline
      $x$ $\sim$ $G$($z$) & & ($n$, 1, 4096)\\
      Conv1-D (stride=4) & (25, 2, 64) & ($n$, 64, 1024)\\
      LReLU($\alpha=0.2$) & & ($n$, 64, 1024)\\
      Conv1-D (stride=4) & (25, 64, 128) & ($n$, 128, 256)\\
      LReLU($\alpha=0.2$) & & ($n$, 128, 256)\\
      Conv1-D (stride=4) & (25, 128, 256) & ($n$, 256, 64)\\
      LReLU($\alpha=0.2$) & & ($n$, 256, 64)\\      
      Conv1-D (stride=4) & (25, 256, 512) & ($n$, 512, 16)\\
      LReLU($\alpha=0.2$) & & ($n$, 512, 16)\\
      Conv1-D (stride=4) & (25, 512, 1024) & ($n$, 1024, 4)\\
      LReLU($\alpha=0.2$) & & ($n$, 1024, 4)\\
      Reshape & & ($n$, 4096) \\
      Dense & (1024, 1) & ($n$, 1) \\
      \hline
    \end{tabular}
\end{table}

\subsection{Short-Time Fourier Transform GAN (STFT-GAN)}

GANs based on STFT representations have been proposed for audio generation, e.g., see \cite{Engel2019, Nistal2021}.  Here, we used a similar model, denoted STFT-GAN, based on the DCGAN architecture \cite{Radford2015}.

The discrete STFT for a real-valued time series is calculated by dividing the time series into shorter segments of equal length, multiplying by a window function, and then calculating the one-sided discrete Fourier transform (DFT) on each segment \cite{Smith2011, Mallat2009}.  Unless stated otherwise, we used a Hann window of length $128$ with $50$\% segment overlap, which for a $4096$ length time series produces a (one-sided) STFT with dimensions of $65 \times 65$.  In this case, the constant-overlap-add (COLA) constraint is satisfied, and the STFT can be inverted to obtain a time series of the original length \cite{Smith2011}.  

Tables~\ref{tab:STFT-GAN-gen} and \ref{tab:STFT-GAN-disc} outline the architectures for the STFT-GAN generator and discriminator, respectively, which are composed of five $2$-D convolutional layers with $5 \times 5$ kernels.  The notation in the tables is similar to that used previously, with Conv\,2-D and Transpose Conv\,2-D indicating two-dimensional convolutional and transposed convolutional layers, and $n$ denoting the batch size.  Because the discrete STFT is complex-valued, each STFT entry requires two channels, corresponding to the real and imaginary parts, respectively.  

\begin{table}[p]
  \centering
    \caption{STFT-GAN generator architecture}
    \label{tab:STFT-GAN-gen}
    \begin{tabular}{|l|c|r|} 
    \hline
      \textbf{Operation} & \textbf{Filter Shape} & \textbf{Output Shape}\\
      \hline
      $z$ $\sim$ Uniform(-1, 1) &  & ($n$, 100) \\
      Dense & (100, 4096) & ($n$, 4096) \\
      Reshape & & ($n$, 1024, 2, 2) \\
      ReLU & & ($n$, 1024, 2, 2)\\
      Transpose Conv2-D (stride=2) & (5, 5, 1024, 512) & ($n$, 512, 4, 4)\\
      ReLU & & ($n$, 512, 4, 4)\\
      Transpose Conv2-D (stride=2) & (5, 5, 512, 256) & ($n$, 256, 8, 8)\\
      ReLU & & ($n$, 256, 8, 8)\\
      Transpose Conv2-D (stride=2) & (5, 5, 256, 128) & ($n$, 128, 16, 16)\\
      ReLU & & ($n$, 128, 16, 16)\\
      Transpose Conv2-D (stride=2) & (5, 5, 128, 64) & ($n$, 64, 32, 32)\\
      ReLU & & ($n$, 128, 32, 32)\\
      Transpose Conv2-D (stride=2) & (5, 5, 64, 2) & ($n$, 2, 65, 65)\\
      Tanh & & ($n$, 2, 65, 65)\\
      \hline
    \end{tabular}
\end{table}

\begin{table}[p]
  \centering
    \caption{STFT-GAN discriminator architecture}
    \label{tab:STFT-GAN-disc}
    \begin{tabular}{|l|c|r|} 
    \hline
      \textbf{Operation} & \textbf{Filter Size} & \textbf{Output Shape}\\
      \hline
      $x$ $\sim$ $G$($z$) & & ($n$, 2, 65, 65)\\
      Conv2-D (stride=2) & (5, 5, 2, 64) & ($n$, 64, 32, 32)\\
      LReLU($\alpha=0.2$) & & ($n$, 64, 32, 32)\\
      Conv2-D (stride=2) & (5, 5, 64, 128) & ($n$, 128, 16, 16)\\
      LReLU($\alpha=0.2$) & & ($n$, 128, 16, 16)\\
      Conv2-D (stride=2) & (5, 5, 128, 256) & ($n$, 256, 8, 8)\\
      LReLU($\alpha=0.2$) & & ($n$, 256, 8, 8)\\
      Conv2-D (stride=2) & (5, 5, 256, 512) & ($n$, 512, 4, 4)\\
      LReLU($\alpha=0.2$) & & ($n$, 512, 4, 4)\\
      Conv2-D (stride=2) & (5, 5, 512, 1024) & ($n$, 1024, 2, 2)\\
      LReLU($\alpha=0.2$) & & ($n$, 1024, 2, 2)\\
      Reshape & & ($n$, 4096) \\
      Dense & (4096, 1) & ($n$, 1) \\
      \hline
    \end{tabular}
\end{table}

\section{Training and Implementation}
\label{sec:training}

\subsection{Baseline Implementation}
Following the original WaveGAN training implementation \cite{Donahue2019}, both models were trained using Wasserstein GAN loss with gradient penalty \cite{Gulrajani2017} and the \textsc{Adam} optimizer \cite{Kingma2015}, with WaveGAN using hyperparameter settings of $\alpha=10^{-4}$, $\beta_1 = 0.5$, and $\beta_2 = 0.9$ for the learning rate and moment decay rates, respectively.  STFT-GAN differs by setting $\beta_1 = 0$, which was recommended by Gulrajani et al. \cite{Gulrajani2017}. 

Consistent with the original Wasserstein GAN implementation \cite{Arjovsky2017}, WaveGAN was trained with an imbalanced discriminator-generator update rule, where the discriminator weights were updated five times for each generator update.  In contrast, STFT-GAN was trained with a balanced discriminator-generator update rule, where the discriminator weights were updated once for each generator update.  The balanced update rule for STFT-GAN was selected based on limited tests carried out for a prior study \cite{Sklar2021}, where we found that balanced updates yielded improved convergence for STFT-GAN.     

Each model was trained with a target data set of size $2^{14} = 16\,384$, for $500$ epochs with a batch size of $128$.  These parameter values were found to be sufficient for convergent training across all experiments.  The data accompanying this paper \cite{noiseGAN-data} include GAN training history files as well as plots of GAN loss and discriminator output during training.  

Prior to training, target distribution training sets were scaled using feature min-max scaling, which scales each feature, i.e., time sample or pixel value\footnote{The real and imaginary channels of complex-valued STFT pixels were scaled separately.}, to the interval $[-1, 1]$, the range of the hyperbolic tangent output activation of the generator. Specifically, minimum and maximum values of each feature were estimated over the training set of size $16\,384$.  Because the generator's output activation is a hyperbolic tangent function, the raw generated data was in the range [-1, 1].  Raw generated data was rescaled using the inverse feature min-max transformation with the minimum and maximum values estimated from the training set.  Therefore, the range of generated data was restricted to the range of the training dataset.  

\subsection{Quantile Data Transformation for Impulsive Noise}

As we will see later, the impulsive noise types were particularly challenging for our baseline GAN models.  Consequently, for the impulsive noise types, we also investigated replacing the feature min-max scaling of the target data with a quantile transformation \cite[Sec. 7.4.1]{DasGupta2010} applied independently to each channel to make the data approximately follow a standard normal distribution.  The motivations for this transformation were twofold: (1) it ensured that the distribution of each channel was unimodal with ``light'' tails \cite{Nair2022}, and (2) it effectively limited the impact of outliers.    

We implemented the quantile transformation using the ``quantile\_transform'' method in the scikit-learn Python library \cite{scikit-learn2011}.  This method is based on the formula $Y = F_Y^{-1}(F_X(X))$, where $X$ is an input random variable with continuous cumulative distribution function (CDF) $F_X(x)$, and $Y$ is an output random variable with desired continuous CDF $F_Y(y)$.  In our case, $F_Y(y)$ is the CDF for a standard normal distribution.  The transformation formula follows from the fact that the random variable $F_X(X)$ has a uniform distribution on the interval $[0,1]$ \cite[Sec. 7.4.1]{DasGupta2010}.  In practice, to apply this method to a sample of $X$, $F_X$ is replaced by the empirical CDF.  

The quantile transformation for a given training set was estimated using 1024 uniformly-spaced quantiles for each target distribution channel.  Any data values exceeding the most extreme quantiles were clipped to those values.  For WaveGAN, the transformation was fit directly to the time series values, whereas for STFT-GAN, the transformation was fit on the real and imaginary channels of the target STFT distribution separately.  For both models, a scaled-tanh activation was used at the end of the generator to limit the absolute-maximum value of generated data to the absolute maximum of the target quantile-transformed distribution.  Finally, the inverse quantile transformation was applied to each channel of the generated data to return it to the original range.   

While the quantile transformation method is included in the commonly used scikit-learn Python library, to our knowledge, it has not been previously examined as a preprocessing step for GAN training.

\section{Evaluation Methods}

Performance evaluation of generative models, and GANs in particular, is a difficult problem and an active research area.  Recent developments are summarized in two review papers by Borji \cite{Borji2019, Borji2022}.  Two important aspects of generative model quality are fidelity, i.e., the degree of realism in generated samples, and diversity, i.e., how well generated samples capture the full range of variation of the target distribution \cite{Borji2019, Naeem2020}.  

We assessed fidelity and diversity using general-purpose metrics introduced by Naeem et al. \cite{Naeem2020}, described below.  In addition, we further evaluated generative fidelity in terms of median power spectral density (PSD) and characteristic parameters for each noise type. Evaluations for each noise type were conducted using test sets of size $4096$ from the target and generated time series distributions.  In particular, the target distribution test sets were synthesized independently from the training sets.  

\subsection{Density and Coverage Metrics}

In an effort to to address shortcomings of other evaluation measures, Naeem et al. \cite{Naeem2020} proposed general-purpose metrics named \emph{density} and \emph{coverage} to assess generative model fidelity and diversity, respectively. 

Suppose that a suitable metric space for the data is identified, and denote test samples of real (target) data as $X_1, X_2, \ldots, X_N$ and fake (generated) data as $Y_1, Y_2, \ldots, Y_M$.   For a given real data sample, $X_i$, let $\text{NND}_k(X_i)$ be the distance from $X_i$ to the $k$th nearest neighbor among the real data sample excluding itself, and let $B(x, r)$ denote the ball centered at $x$ with radius $r$.  Also, let $\mathcal{I}[S]$ be the indicator function that equals one if the proposition $S$ is true and zero otherwise.

For a given fake sample, $Y_j$, Naeem at al. \cite{Naeem2020} define density as the expected number of real sample neighborhoods that contain $Y_j$ divided by the expected number of such neighborhoods when the target and generated distributions are the same.  Namely, for a given test sample, Naeem at al. propose the estimator\footnote{We use ``hat'' notation to distinguish point estimators from the quantity being estimated.}  
\begin{equation}
\widehat{\text{density}} = \frac{1}{kM} \sum_{j=1}^M \sum_{i=1}^N \mathcal{I}\left[Y_j \in B(X_i, \text{NND}_k(X_i))\right],
\label{eq:density}
\end{equation}
where division by $kM$ ensures that $E[\widehat{\text{density}}] = 1$ when the real and fake distributions are the same \cite[Lemma~1]{Naeem2020}.  Note that while density is always greater than or equal to zero, it may be greater than one, depending on the density of real data around the fake data.  Density values close to one indicate excellent generative model fidelity.  On the other hand, values near zero indicate poor fidelity.  Naeem et al. \cite{Naeem2020} do not comment on how to interpret density values much larger than one, so additional assessments of generative fidelity are likely needed in that circumstance.  

To evaluate generative diversity, Naeem et al. \cite{Naeem2020} define coverage as the fraction of real samples whose neighborhoods contain at least one fake sample.  For a given test sample, Naeem at al. estimate coverage as
\begin{equation}
    \widehat{\text{coverage}} = \frac{1}{N} \sum_{i=1}^N \mathcal{I}\left[ \exists j~\text{s.t.}~Y_j \in B(X_i, \text{NND}_k(X_i)) \right].
    \label{eq:coverage}
\end{equation}
Because coverage is essentially the probability that a real sample is ``close'' to a fake sample, it is bounded between zero and one. Coverage values close to one indicate good generative diversity, i.e., generated samples cover the full support of the target distribution.  Conversely, coverage values near zero imply poor generative diversity, which may result from mode collapse (a.k.a. mode dropping), as demonstrated by Naeem at al. \cite{Naeem2020}.

Under the condition that the real and fake distributions are identical, Naeem at al. show that \cite[Lemma~2]{Naeem2020}
\begin{equation}
    E[\widehat{\text{coverage}}] = 1 - \frac{(N-1) \cdots (N-k)}{(M+N-1)\cdots (M+N-k)}.
\end{equation}
Moreover, they propose that the hyperparameter, $k$, should be selected to ensure that the expected value of the coverage estimator is close to one when the real and fake distributions are the same.  In our evaluations, we used test sets of size $M=N=4096$ and implemented the above density and coverage estimators with $k=10$, implying that $E[\widehat{\text{coverage}}] \approx 0.999$ when the target and generated distributions are identical.  

Application of the above density and coverage metrics requires defining a suitable measure of distance between data points.  We chose to use a normalized version of dynamic time warping (DTW) distance \cite{Berndt1994, Keogh2005}, a widely-used general-purpose distance measure for time series indexing, classification and clustering \cite{Keogh2005, Ding2008, Bagnall2017} that is considered a ``standard'' elastic distance measure in the data mining community \cite{Bagnall2017}.  To compute DTW distances between time series, we used the ``fast'' methods from the `dtaidistance' python package \cite{DTIADISTANCE2022}, setting the maximal warping window size to $32$.  The window size parameter was selected based on computational feasibility considerations and limited preliminary experiments.  To obtain a robust distance measure that was insensitive to data scaling, each time series was first normalized by its maximum absolute value prior to DTW estimation.  Estimated DTW distances were then normalized by the window size to ensure values between 0 and 1.  

Normalized DTW distances were computed as described above between each target time series as well as between each target and generated time series in the test sets of size $4096$.  Subsequently, density and coverage were estimated using Eq. (\ref{eq:density}) and Eq. (\ref{eq:coverage}), respectively, with $k=10$.  

Approximate 95\% confidence intervals for the density metric were estimated using the percentile bootstrap method \cite{Efron1994}, where bootstrap resampling with replacement was performed over the generated test sample $10\,000$ times.  Limited experiments indicated that additionally bootstrapping over the target distribution sample resulted in bootstrap estimates that were uniformly lower than the original point estimate, so bootstrap resampling was therefore restricted to the generated sample only.  Approximate 95\% confidence intervals for the coverage metric were estimated using the classical Wilson score method for a binomial proportion \cite{Agresti1998}.    

To our knowledge, the combination of the density and coverage metrics above with DTW distance for time series, as well as the procedures for confidence intervals, have not been previously proposed and are novel.  

\subsection{Power Spectral Density}
\label{sec:PSD_estimation}

The median PSD for each test set was estimated with the multitaper method, a versatile nonparametric approach \cite{Thomson1982,Percival2020}.  Specifically, we used the implementation in the Python `Spectrum' package \cite{Cokelaer2017} with the time half-bandwidth parameter set to $NW=4$, the first $k=7$ Slepian sequences, the FFT length set to $4096$, and the fast `eigen' method for result weighting.  These parameter choices are typical and were found to yield consistent results.  After applying the multitaper method to estimate the PSD for each time series in the test set, we calculated the median value in each frequency bin.  The uncertainties in the median PSD estimate across the test set were negligible in the context of our evaluations.  This procedure was carried out on both target and generated distributions across all noise types.    

Denote the one-sided median PSDs for the target and generated distributions as $P_t(f_d)$ and $P_g(f_d)$, respectively, where $f_d \in [0,0.5]$ is normalized digital frequency with units of cycles per sample.  To evaluate the faithfulness of $P_g$ relative to $P_t$, we used a one-sided version of Georgiou's ``geodesic distance'' for power spectra \cite{Georgiou2007}, defined as\footnote{Consistent with Georgiou's definition, the integrals were normalized by the length of the integration interval.}
\begin{equation}
    d_g(P_g,P_t) = 
    \sqrt{\int_0^{0.5}\left( \log \frac{P_g(f_d)}{P_t(f_d)}\right)^2 \frac{df_d}{0.5} - \left(\int_0^{0.5}\log \frac{P_g(f_d)}{P_t(f_d)} \frac{df_d}{0.5} \right)^2}.
\label{eq:geodesic_distance}
\end{equation}
In our evaluations, we used a natural logarithm, but the choice of logarithm is arbitrary.  The geodesic distance can be interpreted as the length of a geodesic connecting points on a manifold of PSDs \cite{Georgiou2006}.  Technically, $d_g$ is a pseudo-metric, because it is insensitive to scaling, i.e., $d_g(P_g,P_t) = d_g(P_g,\kappa P_t)$ for any $\kappa >0$ \cite{Georgiou2007}.  Because the first term depends on the difference of log-transformed power spectra, it reflects differences in areas of both low and high power spectral density. 
 We estimated the geodisic PSD distance by approximating the above formula on a discrete frequency grid.  

\subsection{Noise Model Parameters}
\label{sec:noise_param_estimation}

For each noise type, except for band-limited thermal noise, we assessed how well the generated time series distribution matched the target distribution in terms of characteristic noise parameters.  Later, boxplots are used to compare distributions of estimated noise parameters for target and generated time series distributions.  Boxplots of parameter estimates for target distributions with known ground truth characterize the inherent bias and variability of the estimators and hence provide a basis for assessing generated data.  

For power law noise distributions, we evaluated the accuracy of the the Hurst index, $H$, using the well-studied ``discrete variations'' method \cite{Istas1997,Coeurjolly2001,Coeurjolly2017} implemented with a second-order difference filter. 

Under the assumption that the shot noise pulse amplitudes follow an exponential distribution, which is true for our target distributions, we assessed the shot noise event rate, $\nu$, using the (apparently novel) estimator
\begin{equation}
    \hat{\nu} = \frac{2\hat{\mu}_X^2 I_2}{\hat{\sigma}_X^2 I_1^2},
\label{eq:sn_event_rate_estimator}
\end{equation}
where $\hat{\mu}_X$ and $\hat{\sigma}_X^2$ are the estimated mean and variance of the shot noise time series, and where $I_1 = \int_{-\infty}^{\infty} p(t)\,dt$ and $I_2 = \int_{-\infty}^{\infty} p(t)^2 \,dt$ are integrals of the known pulse function, $p(t)$; see Table~\ref{tab:pulse_functions}.  A derivation is given in the Appendix.  

For each of the impulsive noise models, we evaluated two characteristic parameters.  Namely, for BG noise, we assessed the impulse probability, $p$, and the scale parameter ratio, $\theta = \sqrt{\sigma_w^2 + \sigma_i^2}/\sigma_w$, which measures the relative dispersion of the mixture components; see equation (\ref{eq:BG_mixture}).  The BG parameters were estimated by fitting a two-component Gaussian mixture model using the iterative expectation maximization method implemented in the scikit-learn Python library \cite{scikit-learn2011}.  To assess $S \alpha S$ noise, we estimated the characteristic exponent, $\alpha$, and the scale parameter, $\gamma$, using the ``fast'' methods of Tsihrintzis and Nikias \cite{Tsihrintzis1996}.  

\section{Experimental Results}
\label{sec:experiments}

\subsection{Band-limited Thermal Noise}
\label{sec:experiments_BLWN}

\begin{figure}[tb!]
\centering
\includegraphics[width=0.4\linewidth]{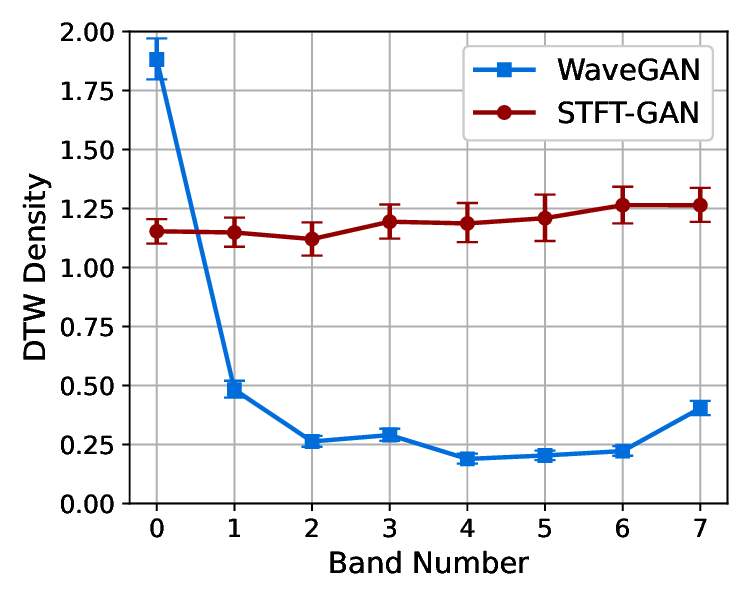}
\includegraphics[width=0.4\linewidth]{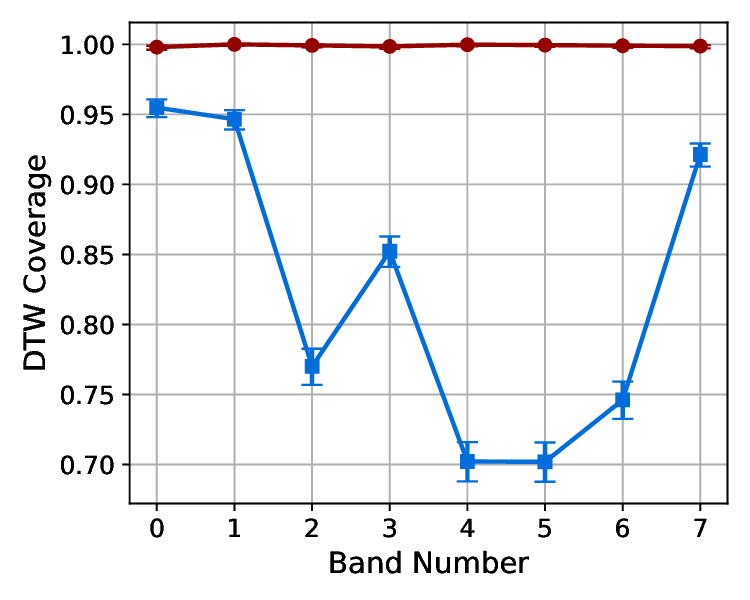}
\caption{DTW density and coverage results for band-limited thermal noise.}
\label{fig:BPWN_DC_results}
\end{figure}

\begin{figure}[tb!]
\centering
\includegraphics[width=0.4\linewidth]{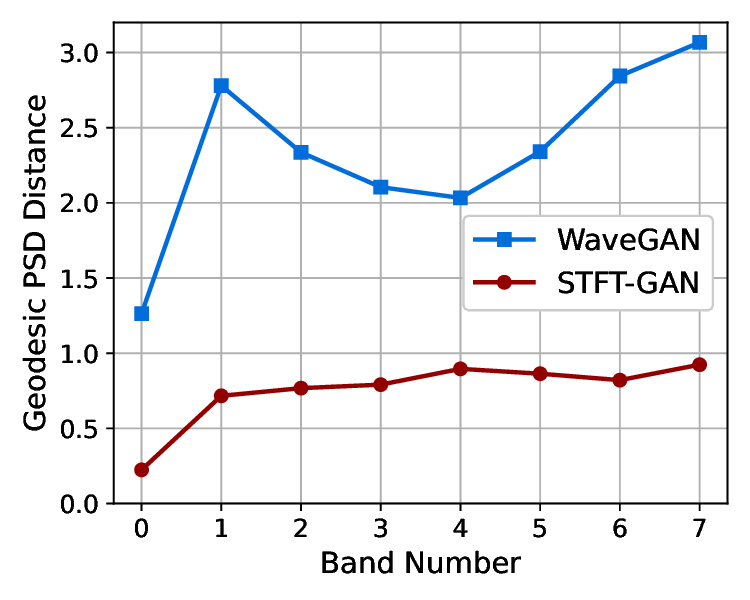}
\includegraphics[width=0.4\linewidth]{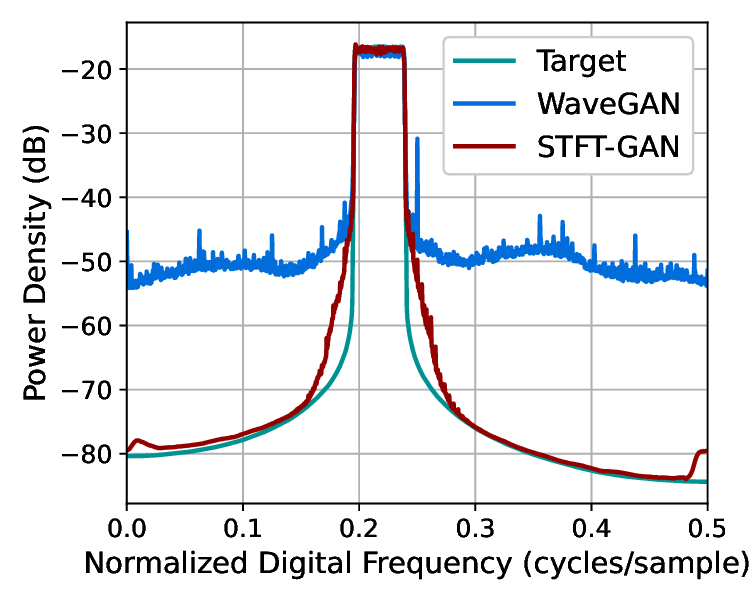}
\caption{PSD results for band-limited thermal noise.  Left: Geodesic PSD distance.  Right: Median PSD comparison for band \#3.}
\label{fig:BPWN_PSD_results}
\end{figure}

The eight digital bandpass filters shown in Figure~\ref{fig:BPfilter_responses} were used to simulate eight target data sets of band-limited thermal noise, where each data set contained noise limited to a single band.  DTW density and coverage results are plotted in Figure~\ref{fig:BPWN_DC_results} and estimated geodesic PSD distance is plotted in Figure~\ref{fig:BPWN_PSD_results} (Left), where the bands are ordered in terms of increasing center frequency.  It is evident that STFT-GAN yielded uniformly better density, coverage, and PSD fidelity than WaveGAN.  Median estimated PSDs for band number 3 are shown in Figure~\ref{fig:BPWN_PSD_results} (Right); other bands are similar.  We see that STFT-GAN more closely tracked the target PSD out-of-band, whereas WaveGAN suffered from a limited dynamic range.

\subsection{Power Law Noise}
\label{sec:pl_noise_results}

\begin{figure}[tb!]
\centering
\includegraphics[width=0.8\linewidth]{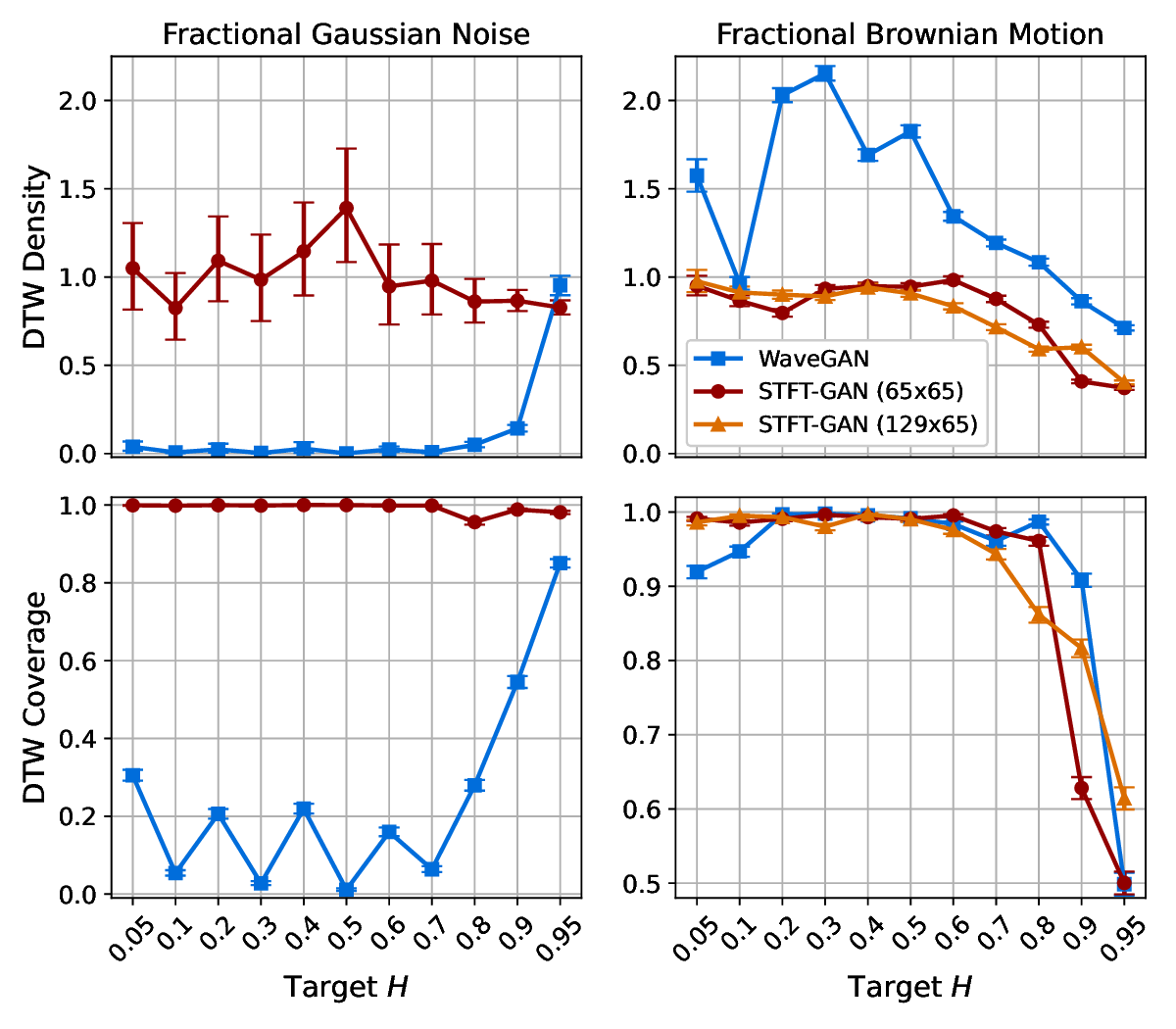}
\caption{DTW density and coverage results for power law noise,}
\label{fig:power_law_noise_DC_results}
\end{figure}

\begin{figure}[tb!]
\centering
\includegraphics[width=0.8\linewidth]{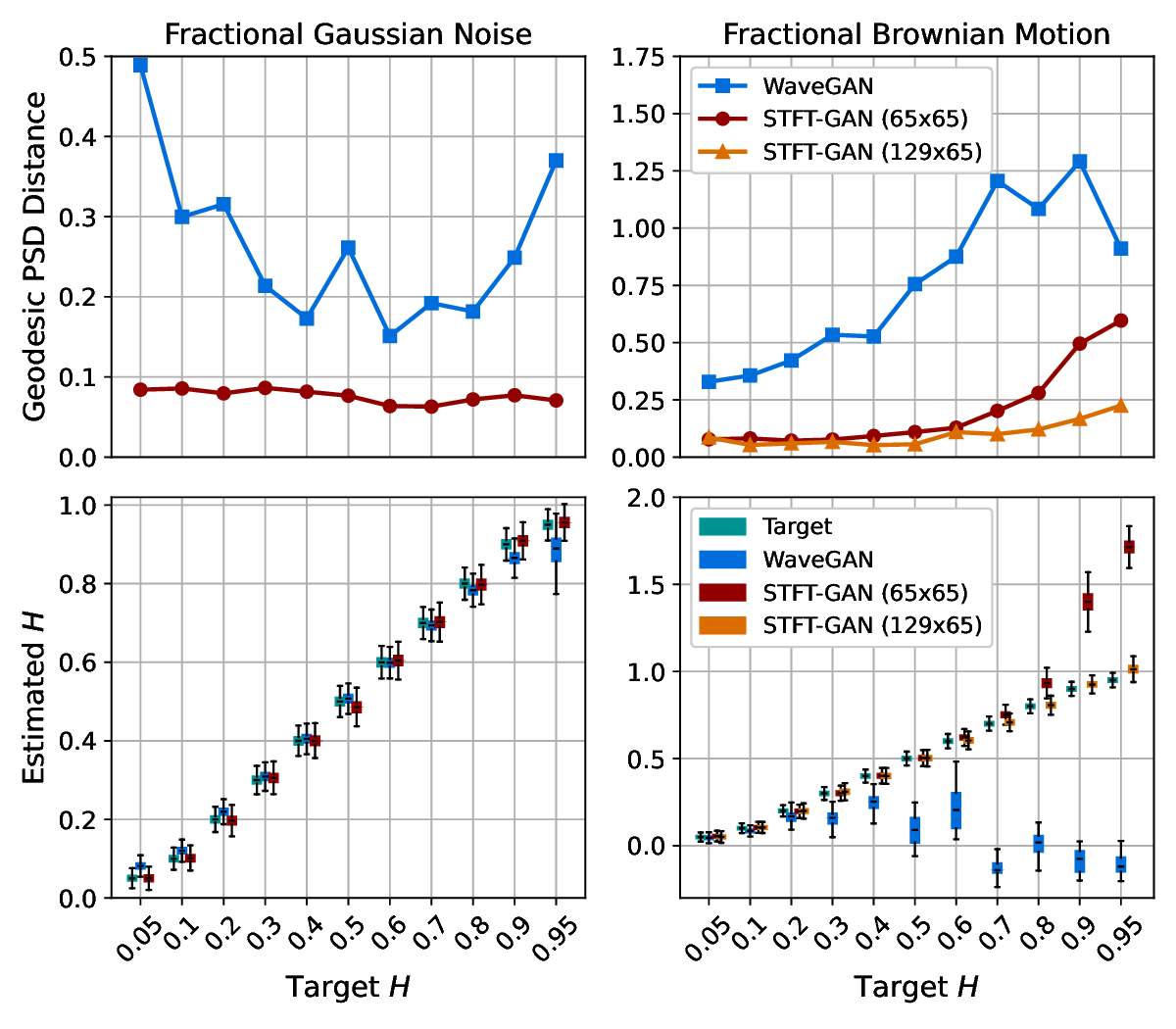}
\caption{Additional fidelity evaluations for power law noise. Top: Geodesic PSD distance plots.  Bottom: Boxplots of estimated Hurst index.}
\label{fig:power_law_noise_results}
\end{figure}

The power law noise models from Section~\ref{sec:power_law_noise_models} were evaluated for target Hurst indices of $H = 0.05$, $0.1$, $0.2$, $0.3$, $0.4$, $0.5$, $0.6$, $0.7$, $0.8$, $0.9$, and $0.95$.  DTW density and coverage results are shown in Figure~\ref{fig:power_law_noise_DC_results}.  PSD distance results and boxplots of estimated Hurst indices are given in Figure~\ref{fig:power_law_noise_results}.  For FGN, STFT-GAN performed well, achieving near-ideal density and coverage as well as excellent median PSD fidelity and Hurst indices.  On the other hand, WaveGAN generally performed poorly on the density and coverage metrics, except for $H=0.95$, and also exhibited inferior median PSD fidelity.   

\begin{figure}[tb!]
\centering
\includegraphics[width=0.5\linewidth]{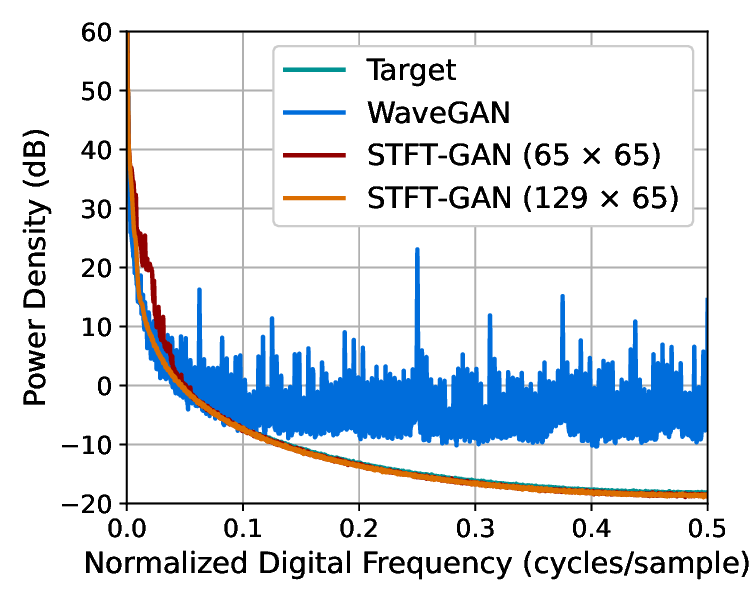}
\caption{Median PSD comparisons for FBM with $H=0.9$.}
\label{fig:FBM_PSD_comparisons}
\end{figure}

For FBM, in addition to the baseline $65 \times 65$ STFT size, we also tested an STFT dimension of $129 \times 65$, resulting from a window segment length of $256$ with $75\%$ overlap.  We denote the baseline and modified models as STFT-GAN ($65 \times 65$) and STFT-GAN ($129 \times 65$), respectively.  Examining DTW density and coverage results, performance was generally excellent except for the largest Hurst indices of $0.9$ and $0.95$, where all models exhibited a drop-off in DTW coverage, indicating inadequate sample diversity.  In terms of median PSD distance and estimated Hurst indices, STFT-GAN ($129 \times 65$), which had higher frequency resolution than the baseline model, achieved superior PSD and Hurst index fidelity over the full parameter range.  Figure~\ref{fig:FBM_PSD_comparisons} compares the median PSDs for the $H=0.9$ case, illustrating better PSD accuracy for STFT-GAN ($129 \times 65$) at low frequencies.  

\begin{figure}[tb!]
\centering
\includegraphics[width=0.6\linewidth]{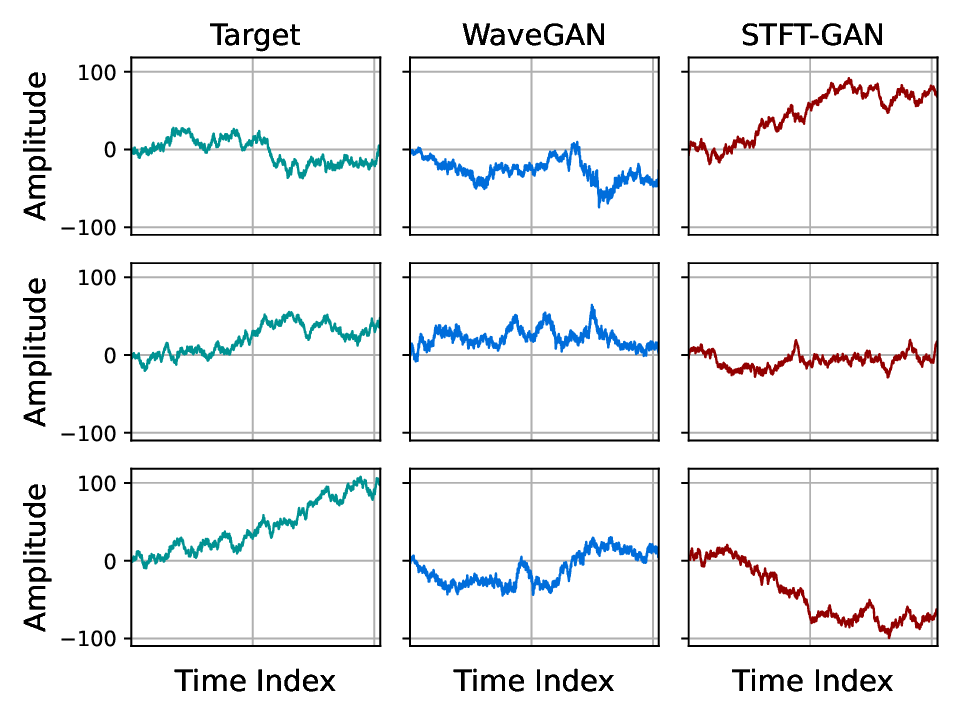} 
\caption{Examples of target and generated time series for FBM with $H=0.5$.  The STFT-GAN results are with an STFT dimension of $129 \times 65$.}
\label{fig:FBM_waveform_examples}
\end{figure}

Example target and generated time series for FBM with $H=0.5$, which corresponds to the classical Brownian motion process, are plotted in Figure~\ref{fig:FBM_waveform_examples}.  Qualitatively, the example generated time series are consistent with a Brownian motion process.  

\clearpage
\subsection{Shot Noise}

\begin{figure}[tb!]
\centering
\includegraphics[width=0.8\linewidth]{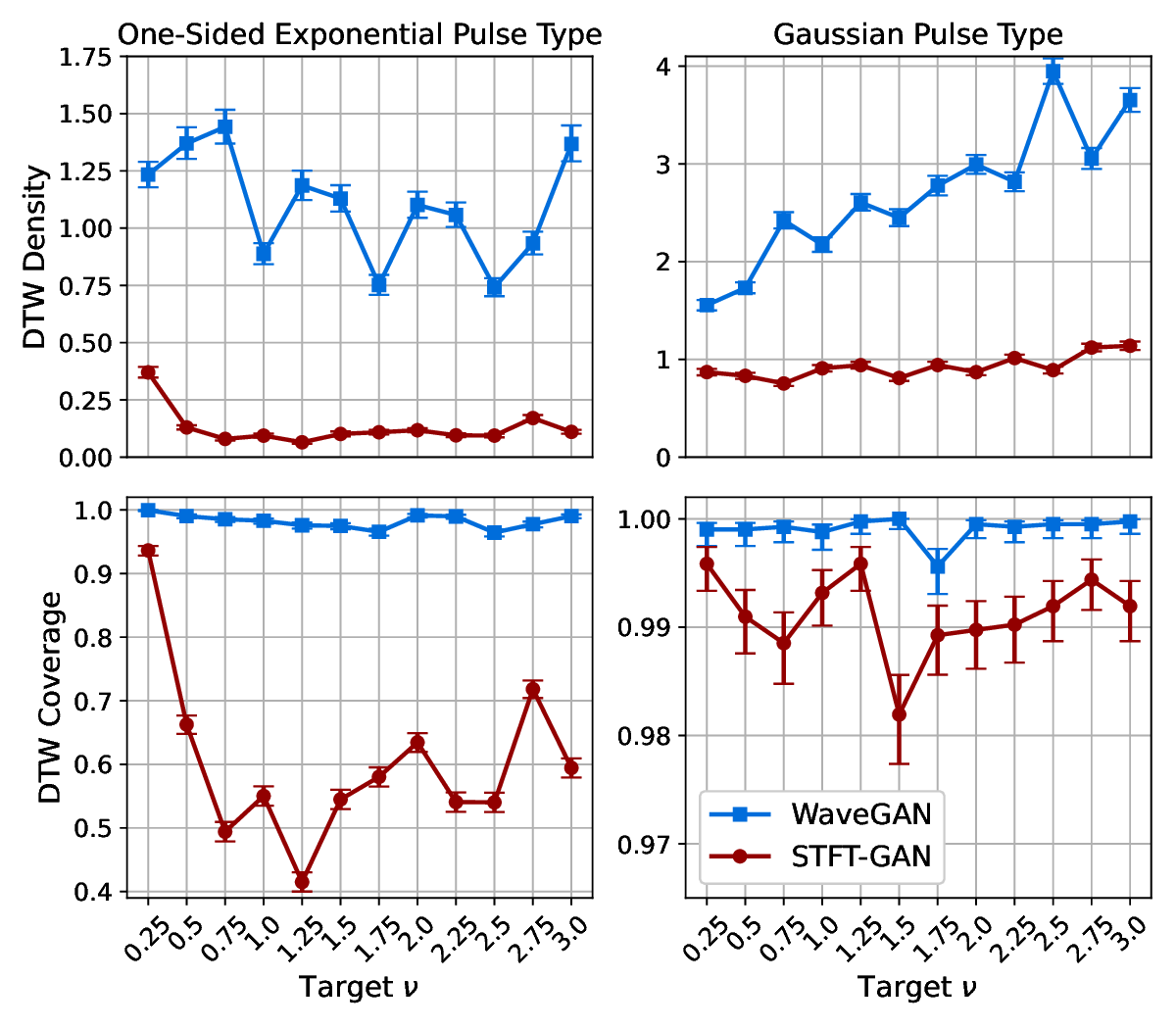}
\caption{DTW density and coverage results for shot noise.}
\label{fig:shot_DC_results}
\end{figure}

\begin{figure}[tb!]
\centering
\includegraphics[width=0.8\linewidth]{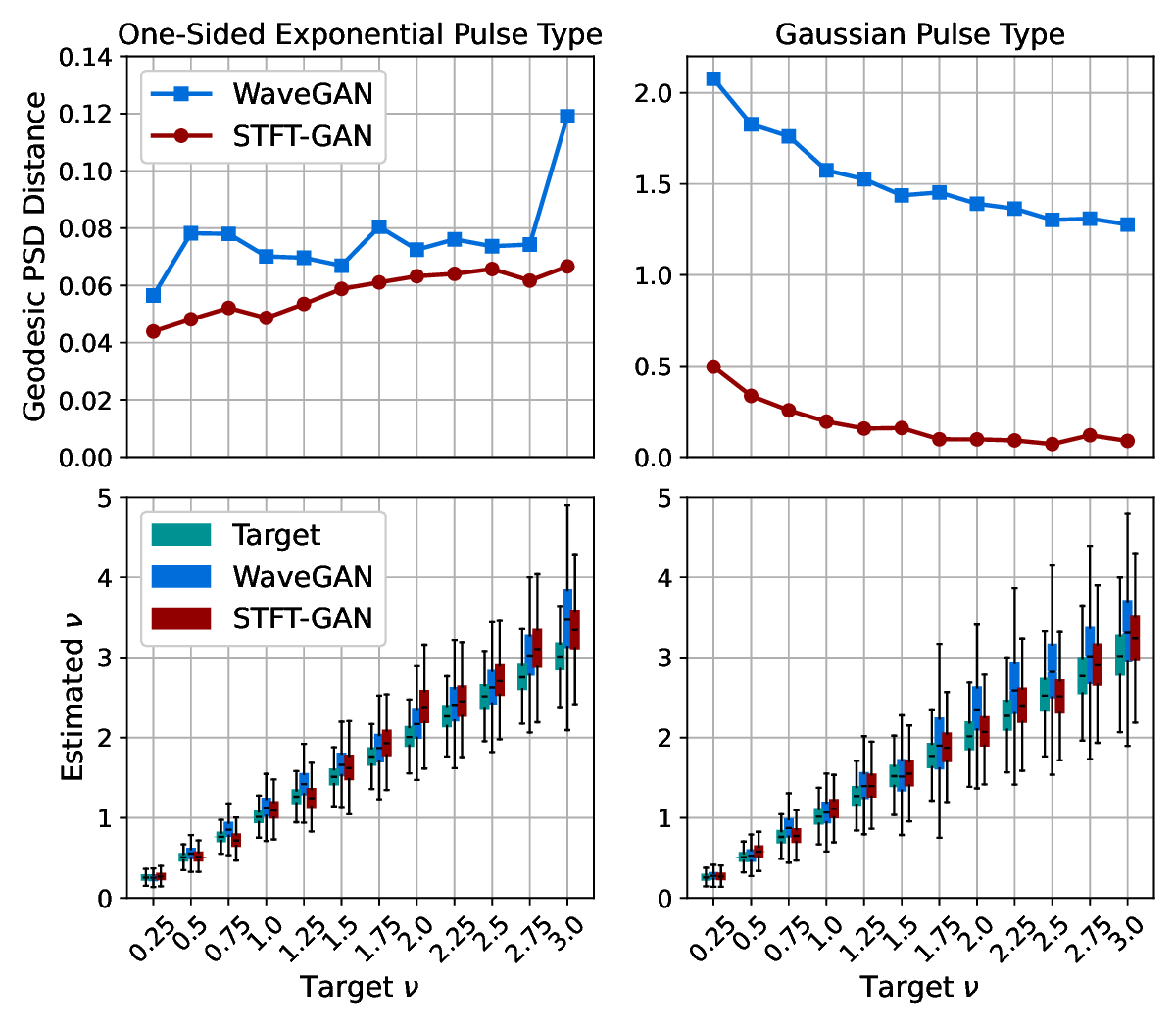}
\caption{Additional fidelity evaluations for shot noise. Top: Geodesic PSD distance plots.  Bottom: Estimated event rate boxplots.}
\label{fig:shot_PSD_event_rate_results}
\end{figure}

Target distributions defined by the shot noise model described in Section~\ref{sec:shot_noise_model} were assessed with the two pulse types in Table~\ref{tab:pulse_functions} for event rates $\nu  = 0.25$, $0.5$, $0.75$, $1.0$, $1.25$, $1.5$, $1.75$, $2.0$, $2.25$, $2.5$, $2.75$, and $3.0$.  DTW density and coverage results are shown in Figure~\ref{fig:shot_DC_results}.   PSD distance results and estimated event rate boxplots are given in Figure~\ref{fig:shot_PSD_event_rate_results}.  

For shot noise with the one-sided exponential pulse type, WaveGAN exhibited very good DTW density and coverage, while STFT-GAN did poorly on those metrics.  Both models had excellent median PSD fidelity and similar event rate performance. 

On the other hand, for shot noise with the smoother Gaussian pulse type, STFT-GAN performed better overall than WaveGAN, with STFT-GAN exhibiting exhibiting excellent DTW density, coverage, and PSD distance.  In this case, WaveGAN achieved worse fidelity as measured both by DTW density and PSD distance. 
 
\begin{figure}[tb!]
\centering
\includegraphics[width=0.49\linewidth]{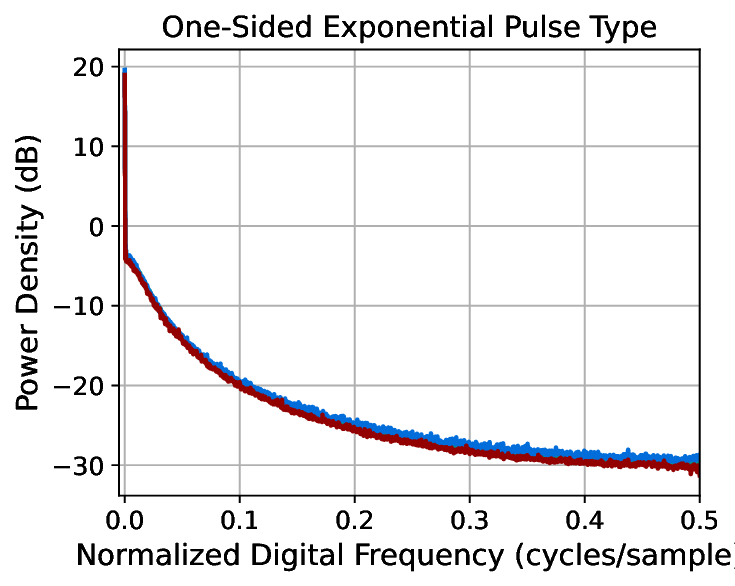} 
\includegraphics[width=0.49\linewidth]{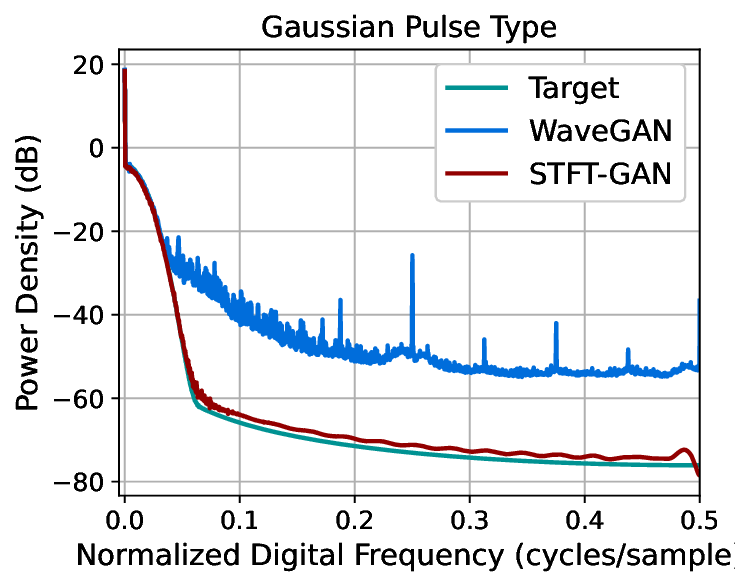} 
\caption{Median PSDs for Shot Noise with event rate $\nu=1$.}
\label{fig:SN_PSD_comparisons}
\end{figure}

Figure~\ref{fig:SN_PSD_comparisons} compares median PSDs for the two pulse types when the target event rate is $\nu=1$.  These plots illustrate the inability of WaveGAN to recover the larger PSD dynamic range for the Gaussian pulse type.    

\begin{figure}[tb!]
\centering
\includegraphics[width=0.6\linewidth]{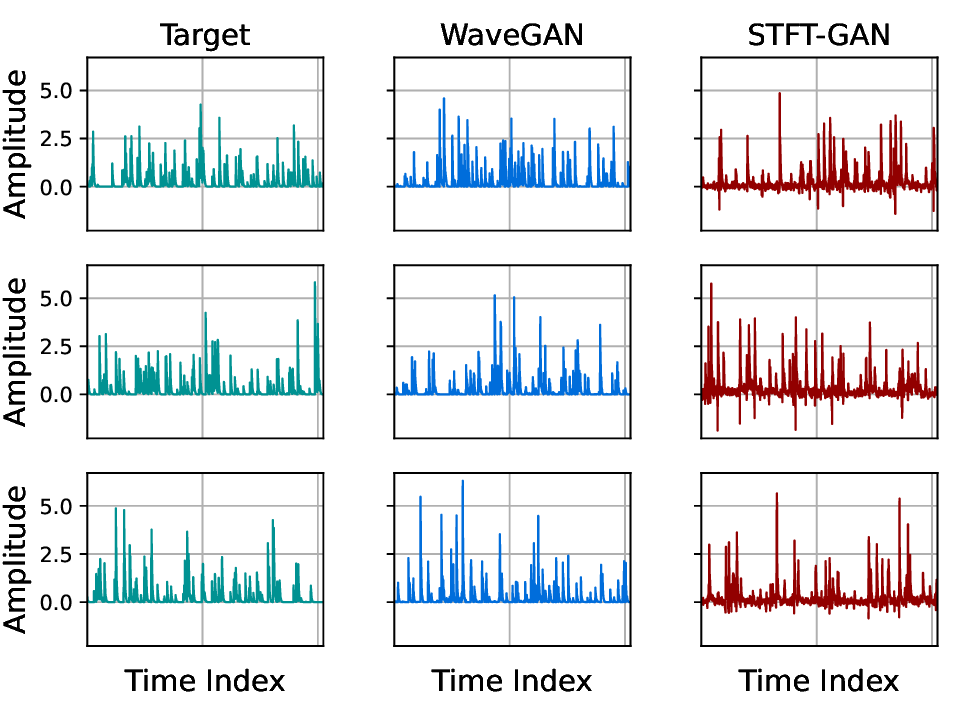} 
\caption{Examples of time series for one-sided exponential shot noise with target event rate $\nu=0.25$.}
\label{fig:SN_waveform_examples}
\end{figure}

Representative example target and generated time series for one-sided exponential shot noise for a target event rate of $\nu=0.25$ are plotted in Figure~\ref{fig:SN_waveform_examples}.  From this figure, it can be seen that WaveGAN correctly learned to generate non-negative shot noise time series, while STFT-GAN generated time series with occasional small negative values.  

\clearpage
\subsection{Impulsive Noise}

Last, the ability of the two GAN models to learn impulsive noise defined by the BG and $S\alpha S$ models described in Section~\ref{sec:impulsive_noise_models} was evaluated.  Specifically, BG noise with $\sigma_w=0.1$ and $\sigma_i=1$, i.e., a scale parameter ratio of $\theta = \sqrt{\sigma_w^2 + \sigma_i^2}/\sigma_w \approx 10.05$, was assessed for impulse probabilities of $p= 0.01$, $0.05$, $0.1$, $0.15$, $0.2$, $0.3$, $0.4$, $0.5$, $0.6$, $0.7$, $0.8$, and $0.9$.  Standard $S \alpha S$ noise with location and scale parameters equal to zero and one, respectively, was evaluated for characteristic exponents $\alpha = 0.5$, $0.6$, $0.7$, $0.8$, $0.9$, $1.0$, $1.1$, $1.2$, $1.3$, $1.4$, and $1.5$.  As described in Section~\ref{sec:training}, the WaveGAN and STFT-GAN models were trained with the two different preprocessing schemes described in Section~\ref{sec:training}: (1) a baseline implementation using feature min-max scaling and (2) an implementation applying a quantile data transformation, which transforms each channel to an approximate standard normal distribution.  

\begin{figure}[tb!]
\centering
\includegraphics[width=0.8\linewidth]{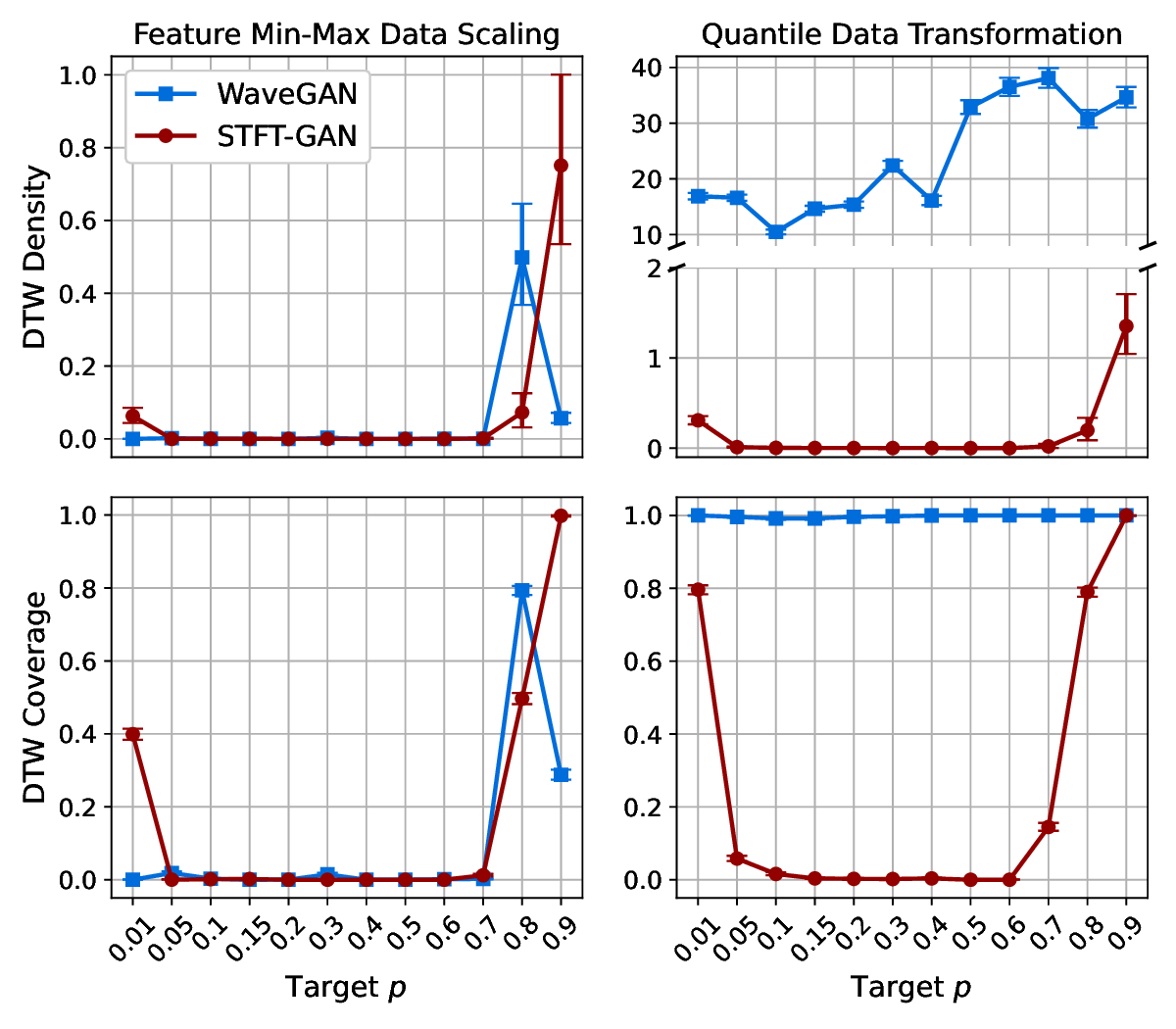}
\caption{DTW density and coverage results for Bernoulli-Gaussian noise.  Note that a broken y-axis is used in the upper right plot to display results for both models.}
\label{fig:bg_dc_results}
\end{figure}

\begin{figure}[tb!]
\centering
\includegraphics[width=0.8\linewidth]{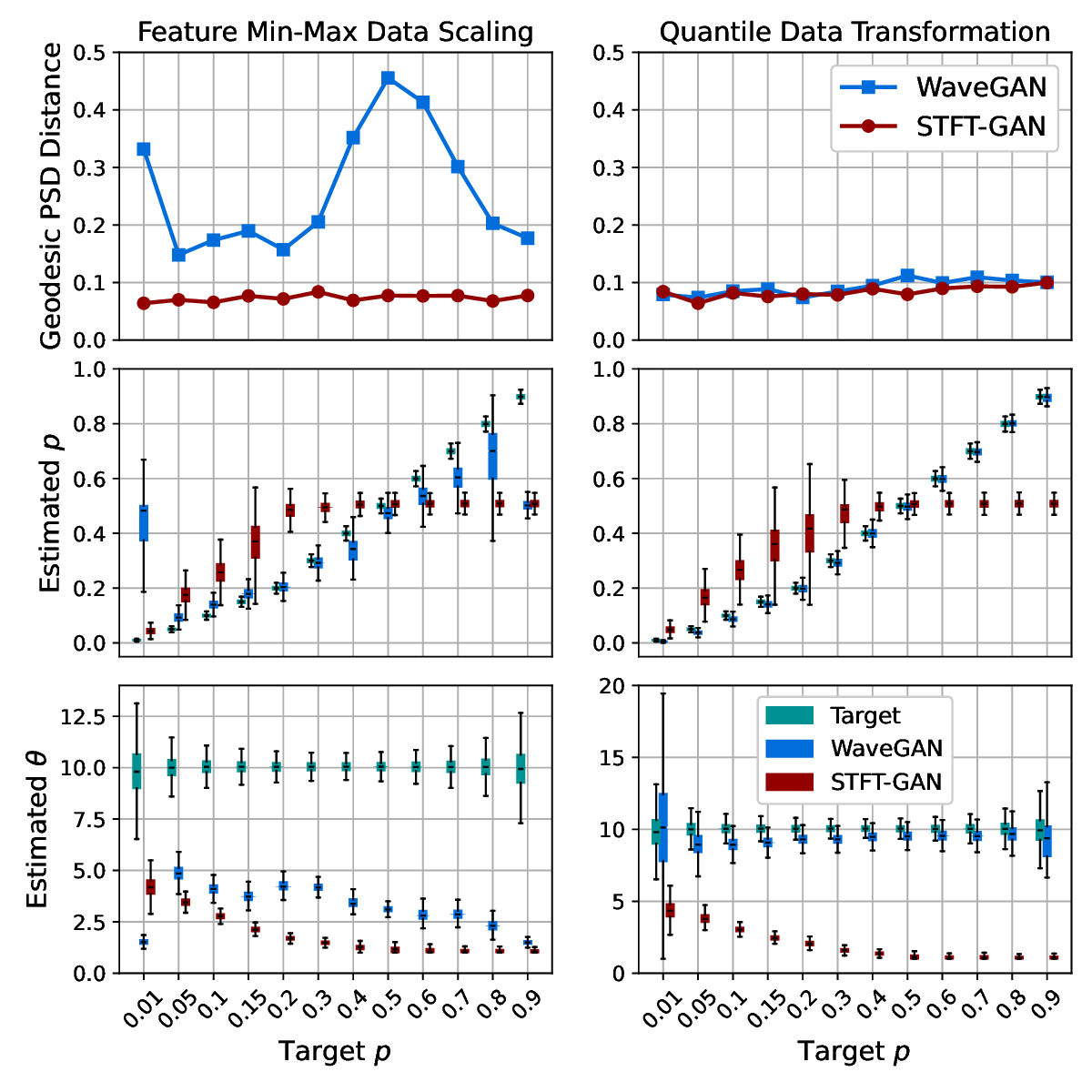}
\caption{Additional fidelity evaluations for Bernoulli-Gaussian noise.  Top: Geodesic PSD distance plots.  Middle: Boxplots of estimated impulse probability.  Bottom: Boxplots of estimated scale parameter ratio.}
\label{fig:bg_extra_fidelity_results}
\end{figure}

Figures~\ref{fig:bg_dc_results} and \ref{fig:bg_extra_fidelity_results} present the aggregate results for Bernoulli-Gaussian noise.  Both GAN models performed poorly with feature min-max scaling, as seen from the DTW density and coverage plots as well as the estimated impulse impulse probability and scale parameter boxplots.  In particular, DTW coverage indicates that both models experienced partial or full mode collapse for all but the largest target impulse probabilities, $p$.  

WaveGAN clearly improved with the quantile data transformation, exhibiting excellent DTW coverage, although the DTW density metric was abnormally large.\footnote{Additional investigations are required to understand the meaning of very large density values.  Naeem at al. \cite{Naeem2020} do not comment on how to interpret this outcome.  Potential explanations include overfitting or a breakdown in the distance measure.} Also, WaveGAN accurately recovered the target impulse probability and scale ratio across most scenarios, except for the extreme $p=0.01$ case, where the dispersion in scale ratio was very large.  By contrast, the quantile data transformation did not appear to improve STFT-GAN performance.  

\begin{figure}[tb!]
\centering
\includegraphics[width=0.6\linewidth]{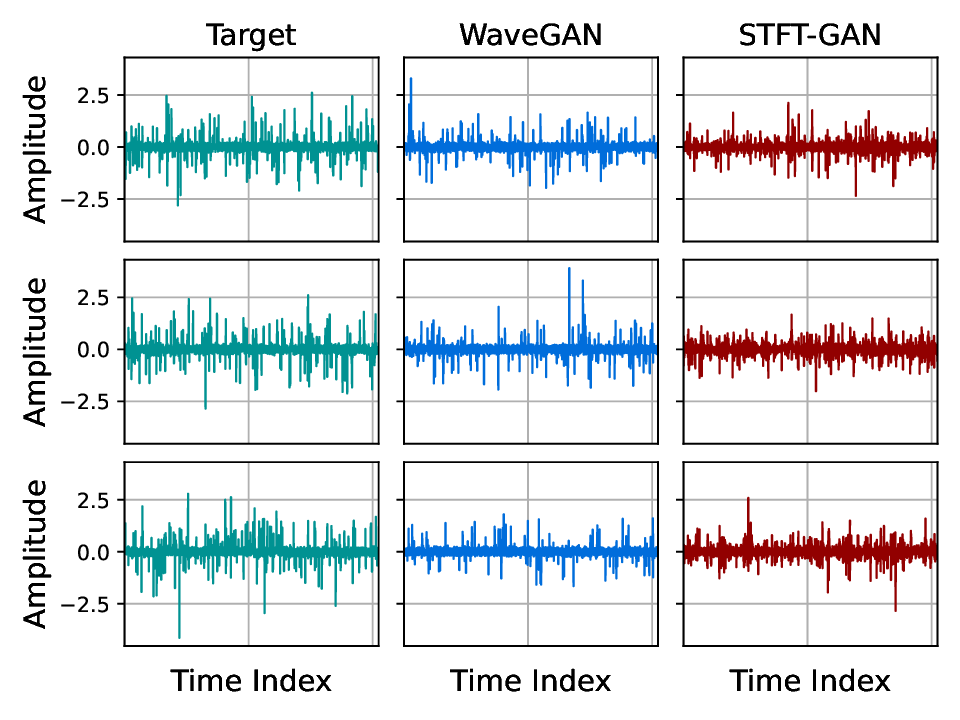} 
\caption{Example time series for GANs trained on Bernoulli-Gaussian noise with quantile data transformation and target impulse probability, $p=0.05$.}
\label{fig:BG_waveform_examples}
\end{figure}

Example target and generated time series for GANs with the quantile data transformation are shown in Figure~\ref{fig:BG_waveform_examples} for the case of $p=0.05$ BG noise.  From these plots, it is evident that STFT-GAN failed to recover the correct background noise level relative to the impulsive component, while WaveGAN better matched the target distribution.  

\clearpage

\begin{figure}[tb!]
\centering
\includegraphics[width=0.8\linewidth]{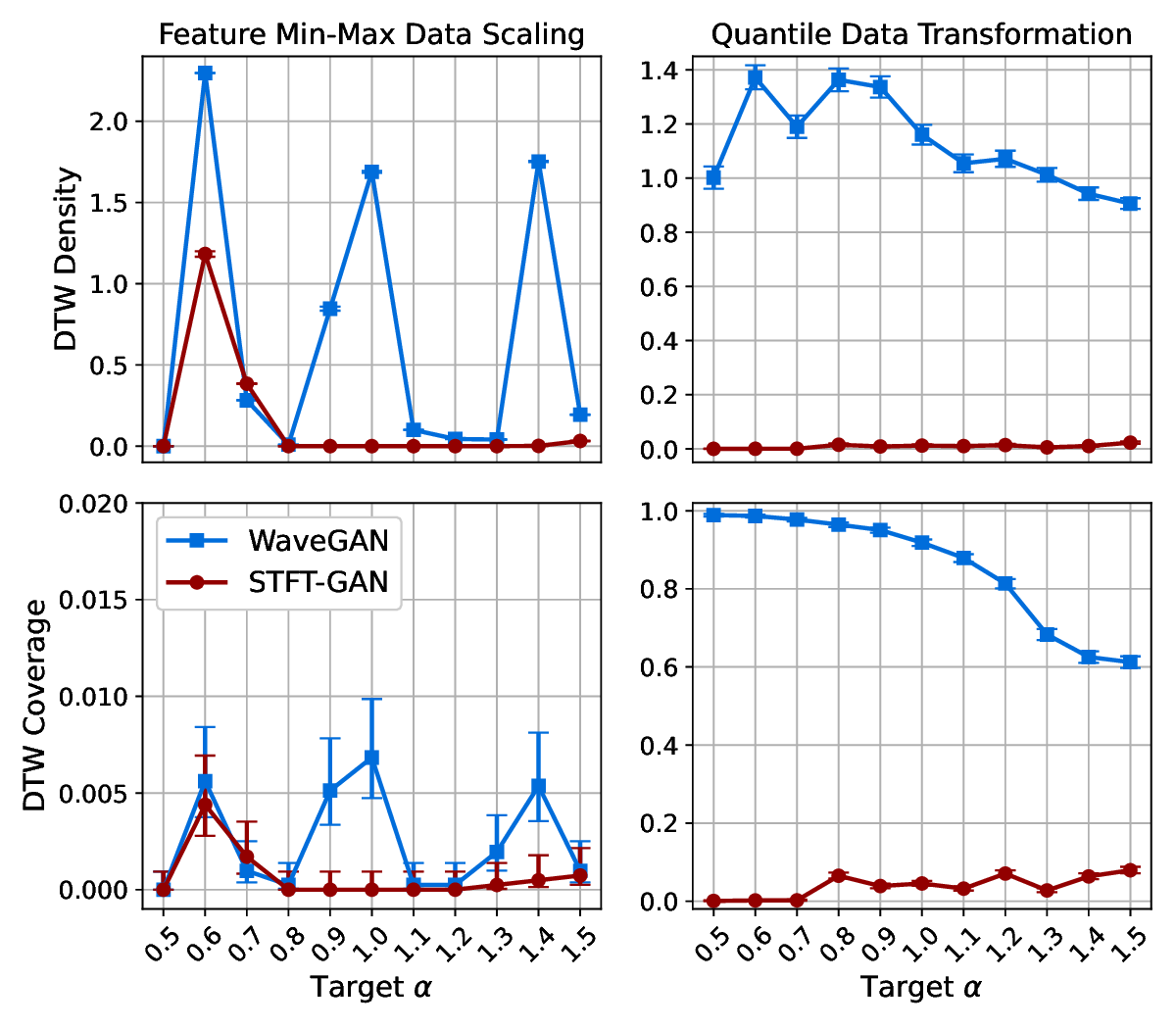}
\caption{DTW density and coverage results for symmetric $\alpha$-stable noise.}
\label{fig:sas_DC_results}
\end{figure}

\begin{figure}[tb!]
\centering
\includegraphics[width=0.8\linewidth]{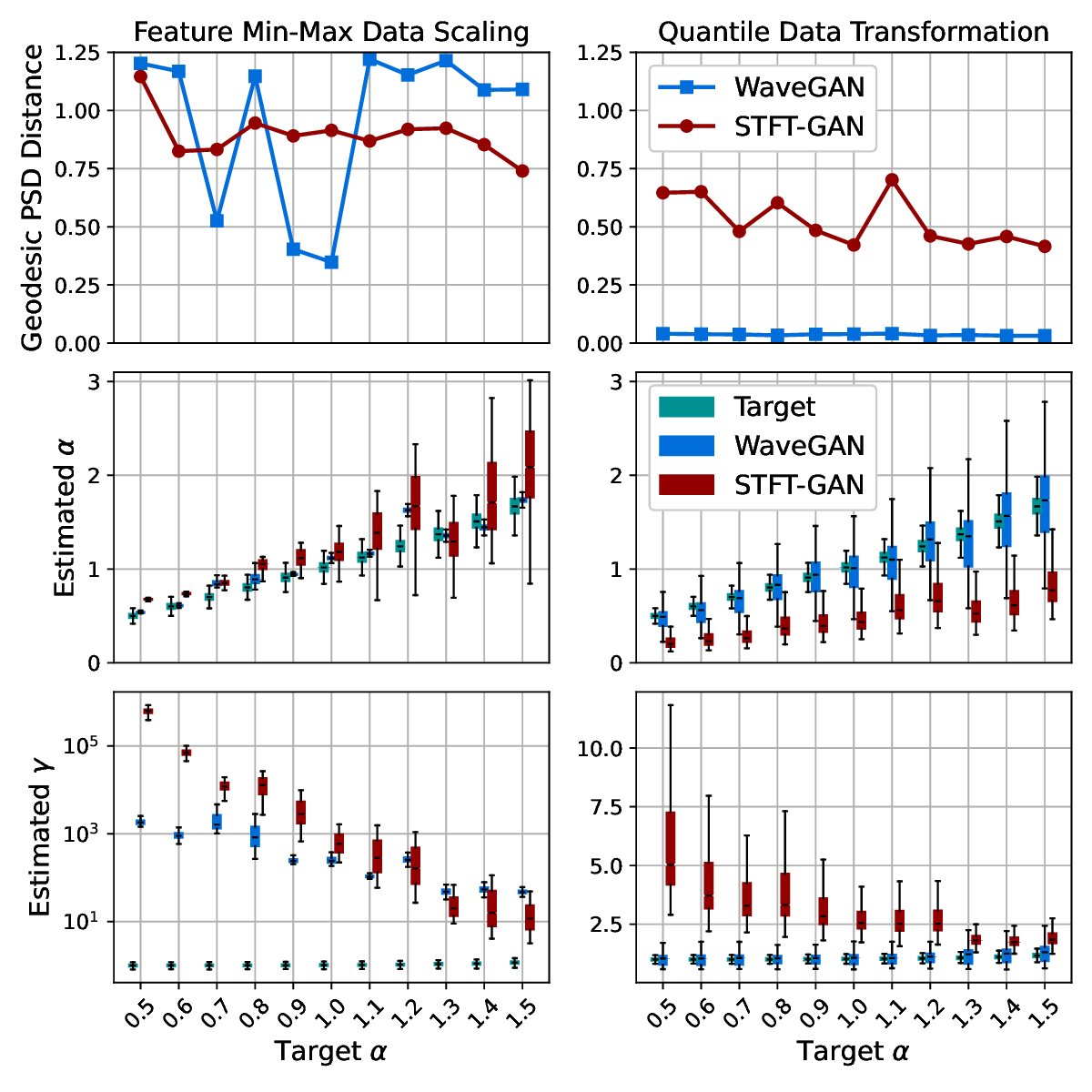}
\caption{Additional fidelity evaluations for symmetric $\alpha$-stable noise.}
\label{fig:sas_extra_fidelity_results}
\end{figure}

Aggregate results for $S\alpha S$ noise are given in Figures~\ref{fig:sas_DC_results} and \ref{fig:sas_extra_fidelity_results}. The GAN models with feature min-max scaling suffered from mode-collapse during training across all tests, as evidenced by the near-zero DTW coverage results.  By contrast, the quantile data transformation preprocessing step enabled WaveGAN to avoid mode-collapse during training, whereas STFT-GAN still suffered from poor diversity, as measured by DTW coverage.  In terms of the fidelity metrics, WaveGAN clearly outperformed STFT-GAN, although the dispersion in the characteristic exponent was unacceptably large.  

\begin{figure}[tb!]
\centering
\includegraphics[width=0.6\linewidth]{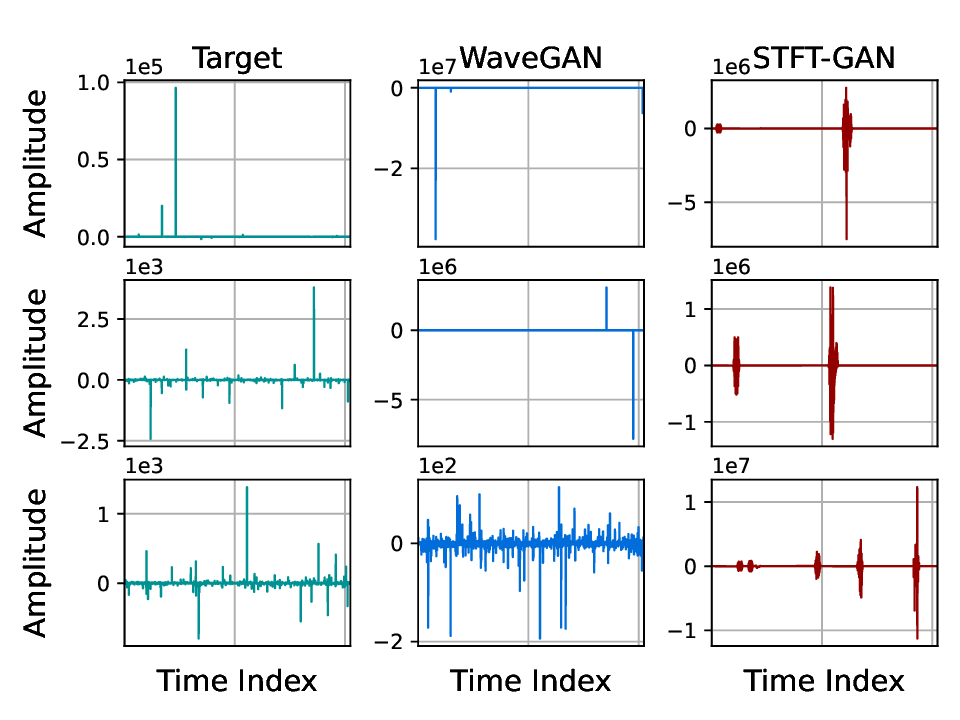} 
\caption{Example time series for GANs trained on symmetric $\alpha$-stable noise with quantile data transformation and target characteristic exponent, $\alpha=1.0$.}
\label{fig:SAS_waveform_examples}
\end{figure}

Example target and generated time series for GANs trained with a quantile data transformation on $S\alpha S$ noise with $\alpha=1.0$ are shown in Figure~\ref{fig:SAS_waveform_examples}.  From these plots, we see that WaveGAN produced short-duration impulses, whereas STFT-GAN produced impulses that were not as localized in time.  These observations are consistent with the PSD distance results.  Further, both models often produced time series with maximum impulse amplitudes that were too large, supporting the finding that they did not consistently recover the target characteristic exponent.   

\clearpage
\section{Discussion and Conclusions}

We examined the ability of two general-purpose GAN models for time series, WaveGAN and STFT-GAN, to faithfully learn several types of noise that frequently arise in physical measurements, signal processing, and communications.  Specifically, we investigated four classes of noise with well-established models: band-limited thermal noise, power law noise, shot noise, and impulsive noise.  In addition, within each noise class, we considered multiple types over a broad range of parameter values.  Performance evaluations examined generative diversity, as measured by DTW coverage, and generative fidelity, as measured by DTW density, median PSD distance, and characteristic model parameters specific to each noise type.  

For most noise types, either the time-domain WaveGAN model or the image-domain STFT-GAN model was more effective.  Namely, STFT-GAN was better at learning a large ($>60$~dB) PSD dynamic range, as evidenced by the results for band-pass thermal noise, power law noise, and shot noise with the Gaussian pulse type.  In addition, the flexibility afforded by the choice of STFT dimensions facilitated improvements in STFT performance, as shown by the FBM evaluations; see Figure~\ref{fig:power_law_noise_results}.  These findings indicate that the time-frequency STFT data representation, which more directly encodes the frequency content of a signal, facilitates learning spectral characteristics.

On the other hand, there was evidence that WaveGAN was more effective at learning time-domain signal characteristics, which are especially important for discontinuous shot and impulsive noise.  Namely, WaveGAN outperformed STFT-GAN for shot noise with one-sided exponential pulses and on both impulsive noise types when a quantile data transformation was applied in preprocessing.  These findings suggest that a hybrid GAN model combining both time-domain and frequency-domain features may be beneficial in some settings.  This is an interesting topic for further research.

GAN frameworks developed on standard image datasets with pixel values in the range [0, 255] typically rescale the data to the range [-1, 1], which corresponds to the range of the hyperbolic tangent activation function that is often the generator output. 
Because most GAN research has focused on standard image datasets consisting of bounded data, finding effective approaches for unbounded target distributions is an important research question.  

Indeed, for the challenging cases of non-Gaussian BG and $S\alpha S$ impulsive noise, we found that GANs were especially sensitive to the data scaling method applied prior to training.  Specifically, STFT-GAN failed badly, regardless of the data scaling method.  On the other hand, WaveGAN required a quantile data transformation, which altered the data distribution to approximate a standard normal distribution, to avoid mode collapse.  These findings highlight the limitations of conventional min-max data scaling on training data and the need for further research into general-purpose GANs that can learn impulsive, non-Gaussian time series.  There has been some recent work specifically focused on GANs for heavy-tailed distributions, e.g., \cite{Zhang2021,Zhou2021,Huster2021}, but it is unclear how well these approaches generalize to other types of distributions.  

The experimental evaluations presented here were necessarily limited in scope.  Specifically, we did not attempt to evaluate all types of time series GANs or to propose a single GAN architecture that works optimally for all noise types.  Moreover, we did not consider complex-valued noise, superpositions of multiple noise types, or time series with a deterministic component.  Evaluating the ability of deep generative models to learn these categories of random processes is of high interest for future studies.  

Our performance evaluations demonstrated the value in estimating multiple types of GAN performance measures, including both general-purpose and data-specific metrics that are easier to interpret.  In particular, our evaluations showed the utility of the general-purpose density and coverage metrics based on DTW distance for time series.  The development and study of general-purpose generative model evaluation measures, particularly for time series, remains an important research topic.  

In conclusion, our findings demonstrate that general-purpose time series GANs based on commonly-used deep convolutional architectures are capable of accurately learning many types of classical noise models, including Gaussian and non-Gaussian distributions, as well as stationary and non-stationary random processes.  These results give further evidence that GANs are a very promising class of generative models for blindly learning a wide variety of time series distributions.  Moreover, our battery of tests with classical random processes provides a useful benchmark to aid further development of deep generative models for time series.

\appendix
\section*{Event Rate Estimation for Shot Noise}
We derive the event rate estimator given in equation (\ref{eq:sn_event_rate_estimator}) for the shot noise model of Section~\ref{sec:shot_noise_model} under an exponential distribution for the pulse amplitudes.  From classical results for filtered Poisson processes \cite{Racicot1971,Howard2016}, the mean and variance of of the shot noise process defined by Eq. (\ref{eq:shot_noise_model}) are $\mu_{X} = \nu E[A_n] I_1$ and $\sigma_X^2 = \nu E[A_n]^2 I_2$, respectively, where $E[\cdot]$ denotes mathematical expectation, $I_1 = \int_{-\infty}^{\infty} p(t)\,dt$, and $I_2 = \int_{-\infty}^{\infty} p(t)^2 \,dt$.  Table~\ref{tab:pulse_functions} lists $I_1$ and $I_2$ for the pulse functions considered here.  Now, assuming that $A_n$ follows an exponential distribution with mean $\beta$, it follows that $E[A_n] = \beta$ and $E[A_n^2] = 2\beta^2$.  Thus, $\mu_X = \nu \beta I_1$ and $\sigma_X^2 = \nu (2\beta^2)I_2$.  Solving the $\mu_X$ equation for $\beta$, substituting the result into the $\sigma_X^2$ equation, solving for $\nu$, and replacing $\mu_X$ and $\sigma_X^2$ by their sample estimates yields the estimator in Eq. (\ref{eq:sn_event_rate_estimator}).  For the special case of a one-sided exponential pulse type, the event rate estimator is equivalent to Eq. (29) in \cite{Theodorsen2017} with $\epsilon=0$.  

\section*{Data Availability Statement}
The data that support the findings of this study are openly available \cite{noiseGAN-software, noiseGAN-data}.

\bibliographystyle{IEEEtran}
\bibliography{references}

\end{document}